\documentclass[%
 reprint,
 amsmath,amssymb,
 aps,
]{revtex4-2}

\usepackage{graphicx}
\usepackage{dcolumn}
\usepackage{bm}

\begin{document}

\title{Motion of the charged test particle in the  spinning nonlinear electromagnetic black hole.}

\author{Nora Breton}
\email{nora@fis.cinvestav.mx}
\affiliation{%
 Departamento de F\'isica, Centro de Investigaci\'on y de Estudios Avanzados del I.P.N., Apdo. Postal 14-740, M\'exico City, Mexico.}%

\author{Gustavo Gutierrez-Cano}
\email{ggutierrezcano@ugto.mx}
\affiliation{Departamento de F\'isica, Divisi\'on de Ciencias e Ingenier\'ias, Campus Le\'on, Universidad de Guanajuato, C.P. 37150, Leon, Mexico.}%

\author{Alberto A. Garcia-Diaz}
\email{aagarcia@fis.cinvestav.mx}
\affiliation{%
Departamento de F\'isica, Centro de Investigaci\'on y de Estudios Avanzados del I.P.N., Apdo. Postal 14-740, M\'exico City, Mexico.
}%

\date{\today}

\begin{abstract}
In this  paper the motion of charged and uncharged test particles in the rotating nonlinearly charged black hole  is examined. 
Its asymptotics  can be de Sitter or anti-de Sitter, depending on the value  of the nonlinear parameter; consequently this BH can present one, two or three horizons, the third one being the cosmological  horizon in the de Sitter case.   Angular and radial  test particle  motions are analyzed  and  compared with its linear electromagnetic counterpart, the  Kerr-Newman black hole (KN-BH). Several differences arise with the KN-BH, namely,  the equatorial asymmetry is enhanced by the NLE field and for charged particles the access to one of the poles is forbidden; besides,  a second circular orbit   in the neighborhood of the external horizon appears; the presence of the nonlinear electromagnetic field increses the curvature producing bounded orbits closer to the horizon.
\end{abstract}

\maketitle


\section{Introduction}

The recent improvement in astrophysical observations as well as the direct gravitational wave detection by LIGO \cite{LIGO2016}, Advanced LIGO and Advanced Virgo \cite{ALAV2021} has lead to the assembly of catalogues of black holes that have stimulated the study of test particle motion in the neighborhood of compact objects. Astrophysical compact objects are rotating and therefore in the context of the  Einstein exact solutions there is a great interest in stationary solutions since, within some approximation, they resemble astrophysical black holes (BH) or compact objects.

Moreover, some compact objects can possess strong magnetic fields in their vicinity and these fields can be described by nonlinear electrodynamics (NLE).
Recently was presented, for static spherically symmetric metrics the general exact
solution of Einstein equations coupled to NLE \cite{Garcia_Gustavo2019}, with an
arbitrary structural metric function, which, via a pair of
independent Einstein equations, allows one to derive the single
associated field tensor component $\mathcal{E}$, and the
Lagrangian--Hamiltonian field function $\mathcal{L}$--$\mathcal{H}$,
which determine the entire solution, should it be singular or
regular; also,  a number of Einstein-NLE static solutions have been derived so far, both singular and regular, however the challenge of determining a NLE stationary solution has been elusive. This was accomplished recently and a stationary solution of the coupled Einstein-NLE equations was presented in \cite{AGarcia2022} and \cite{AGarcia2022B}. These first exact solutions describe a  rotating black hole
endowed with mass, angular momentum,  electromagnetic nonlinear parameter and cosmological constant; they fulfil a set of four
generalized ``Maxwell equations'' for the electrodynamics fields
$F_{\mu\nu}$ and $P_{\mu\nu}$ and two independent Einstein--NLE
equations related with the two independent eigenvalues of the NLE
energy--momentum tensor. The NLE is determined by a Lagrangian
function $L (F,G)$ constructed from the two electromagnetic invariants $F$
and $G$.

We do not claim that these solutions represent in a feasible way  real astrophysical objects, but neutron stars and gravastars are characterized by generating strong magnetic fields, then  these kind of exact solutions of the Einstein-NLE equations can give insight in the search of interesting properties of BH  as well as can be useful as test beds of BH numerical simulations.

In this paper we  examine the rotating axisymmetric solution of the Einstein equations coupled with NLE that was recently presented in \cite{AGarcia2022} and with cosmological constant in \cite{AGarcia2022B}.  This solution is characterized by five physical parameters, namely, the gravitational mass $m$, electric and magnetic charges $f_1$, $g_1$ (comprised in $F_0$), the NLE parameter $\beta$ and the angular momentum $a$ . The existence of horizons allows a BH interpretation, however the solution is not asymptotically flat due to the NLE field that renders a de Sitter or  anti-de Sitter asymptotics.
The solution has the Kerr-Newman limit when the NLE parameter vanishes, $\beta=0$, with the Kerr-Newman (KN) electric charge being $Q_e^2= F_0$.
For charged test particles the angular and radial motions  are analyzed  and  compared with its linear electromagnetic counterpart, the  Kerr-Newman black hole (KN-BH);  several differences  arise in the motion of charged test particles, namely,  the equatorial symmetry is not preserved and the access to one of the poles is forbidden and  a second circular orbit  can appear in the neighborhood of the external horizon; the  allowed regions for the  bounded test particle motion  are illustrated in a series of plots varying the parameters of the system.

The paper is organized as follows, in the next section a brief review of NLE is presented.  In  Section III  the  stationary NLE  BH is introduced and its horizons are analyzed.  In  Section IV  the motion equations are derived;  for charged and uncharged test particle  we analyze  trajectories in both,  $\theta$-motion and  $r$-motion. Some comments on birefringence are presented in Section V and  final remarks are given in  Section VI.


\section{Nonlinear electrodynamics}

For completeness we include the basic features of nonlinear electrodynamics (NLE). 
This theory is constructed
from a Lagrangian function $L=L(F,\,G)$ that depends on the
electromagnetic invariant $F$ and pseudo--scalar $G$

\begin{equation}
F=\frac{1}{4}F_{\mu\nu}F^{\mu\nu},\,{{G}}=\frac{1}{4}{}^\star{{F}}_{\mu\nu}F^{\mu\nu},
\end{equation}
where
the  dual field tensor ${}^\star F_{\mu\nu}$ is defined by

\begin{eqnarray}
&&{}^{\star}F_{\mu \nu}:=\frac{1}{2} \sqrt{-g}\epsilon_{\mu\nu\alpha\beta}F^{\alpha \beta},
\, \,{{}^\star{F^{\alpha\beta}}}=-\frac{1}{2\sqrt{-g}} \epsilon^{\alpha\beta\mu\nu}F_{\mu\nu}.\nonumber\\
 \end{eqnarray}

For coupled NLE with Einstein equations  the corresponding energy--momentum tensor $T^{\mu\nu}$ can be derived from the 
variation of the matter Lagrangian $L_M$  with respect to $g_{\mu\nu}$.
In NLE one uses for the Maxwell limit the Lagrangian function
$L_{Max}(NLE; F)=F^{\alpha\beta}F_{\alpha\beta}/4$, instead of the
standard Maxwell Lagrangian function $L_{Max}( F)=-F$, thus, to
obtain the Maxwell limit, one has to use the Lagrangian function
$L_M=-L(NLE;F,G)$. Therefore, accomplishing the variations, one
arrives at
\begin{eqnarray}\label{EnegryT1}
-T^{\mu\nu}&&=L\, g^{\mu\nu}-L_F\,F^{\mu\sigma}
{F^{\nu}}_{\sigma}-L_G{F^{\mu\sigma}}{}^\star{F^{\nu}}_{\sigma}\nonumber\\&&=:L\,
g^{\mu\nu}-F^{\mu\sigma}{P^\nu}_{\sigma},
\end{eqnarray}
where we have introduced  the new field tensor $P_{\mu\nu}$,
which one identifies as the $P_{\mu\nu}$ field tensor of Pleba\'nski
\cite{SalazarGarciaPleb1987},  or the $p^{kl}$--field tensor of
Born--Infeld \cite{BornInfeld34}. $F_{\mu \nu}$ and $P_{\mu \nu}$ are  related to each other by
\begin{eqnarray}
P_{\mu\nu}&=&2\frac{\partial L}{\partial F^{\mu\nu}}=L_F
F_{\mu\nu}+L_{{G}} {}^{\star}{F}_{\mu\nu}, \nonumber\\
F_{\mu\nu}&=&2\frac{\partial H}{\partial P^{\mu\nu}}=H_P
P_{\mu\nu}+H_{{Q}} {}^{\star}{P}_{\mu\nu}.
\end{eqnarray}

To the antisymmetric  field $P_{\mu\nu}$ there are associated its dual field tensor ${}^\star P_{\mu\nu}$ and the
invariants $P$ and $Q$,
\begin{eqnarray}&&{}^{\star}P_{\mu \nu}:=\frac{1}{2} \sqrt{-g}\epsilon_{\mu\nu\alpha\beta}P^{\alpha
\beta},\, \,{{}^\star{P^{\alpha\beta}}}=-\frac{1}{2\sqrt{-g}}
\epsilon^{\alpha\beta\mu\nu}P_{\mu\nu}, \nonumber\\&&
P=\frac{1}{4}P_{\mu\nu}P^{\mu\nu},\,{{Q}}=\frac{1}{4}
{{}^{\star}{P}}_{\mu\nu}P^{\mu\nu}.
 \end{eqnarray}
The structure function $ H(P,Q)$,  associated with the Lagrangian
function $L(F,G)$, can be determined by a Legendre transformation
\begin{equation}
L(F,G)=\frac{1}{2}F_{\mu\nu}P^{\mu\nu}-H(P,Q).
\end{equation}
The electrodynamics is determined through  the ``Faraday--Maxwell''
electromagnetic field equations, which in vacuum are
\begin{eqnarray}
{{}^\star{F^{\mu\nu}}}_{;\nu}=0 \rightarrow {(\sqrt{-g}{}^\star {F^{\mu\nu}})_{,\nu}=0 },\nonumber\\
{P^{\mu\nu}}_{;\nu}=0
\rightarrow{\left[{\sqrt{-g}L_F\,{F^{\mu\nu}}+\sqrt{-g}L_G{{}^\star
{F^{\mu\nu}}}}\right]_{,\nu}=0},
\label{MF_Eqs}
\end{eqnarray}
that can be written by means of a closed 2--form $d\omega=0$,
\begin{eqnarray*}
\omega=\frac{1}{2}\left(F_{\mu\nu}+{{}^\star
P_{\mu\nu}}\right)dx^\mu\,\wedge
dx^\nu=\frac{1}{2}\left(F_{ab}+{{}^\star P_{ab}}\right)e^a
\wedge\,e^b,\
\end{eqnarray*}
since $F_{\mu\nu}$ and ${{}^\star P_{\mu\nu}}$ are curls.

In nonlinear electrodynamics, the energy--momentum tensor
${T^\mu}_{\nu}$, (\ref{EnegryT1}), allows for two different pairs of
eigenvalues $\{\lambda,\lambda,\Lambda^\prime,\Lambda^\prime\}$. One
can show that a similar property,~i.e., two pairs of different
eigenvalues, is shared by the field tensors $F_{\mu\nu}$ and
$P_{\mu\nu}$, although their eigenvalues are different. Moreover, 
NLE- ${T^\mu}_{\nu}$ possesses a non zero trace

\begin{eqnarray}&&\label{EnergyTrace}
-T:=-{T^\mu}_{\mu}=4 (L - L_F\,{F}- L_G\,G).
\end{eqnarray}
On the other hand, taking into account the relation
${F}^{\mu\sigma}{{}^\star F}_{\nu\sigma}=G\delta^\mu_\nu$, one
determines the traceless NLE energy--momentum tensor
$\Upsilon_{\mu\nu}$  to be

\begin{eqnarray}&&\label{Upsilon}
{\Upsilon^\mu}_{\nu}:={T^\mu}_{\nu}-\frac{T}{4}{\delta^\mu}_{\nu}=
L_F({F}^{\mu\sigma}{F}_{\nu\sigma}-\,F{\delta^\mu}_{\nu}).
\end{eqnarray}

The rotating NLE BH is an exact solution of Eqs. \eqref{MF_Eqs} coupled with Einstein Equations, 
$G_{\mu \nu}= \kappa T_{\mu \nu}$, and we present the line element in the next section.


\section{The  stationary NLE BH}

The  NLE BH  reported in  \cite{AGarcia2022}  is a stationary axisymmetric solution of the Einstein equations coupled with NLE,  characterized by five physical parameters, namely, the gravitational mass $m$, electric and magnetic charges $f_1$, $g_1$ (comprised in $F_0$), the NLE  parameter $\beta$ and the angular momentum $a$; the  Kerr-like line element is given by

\begin{eqnarray}
\label{K_like}
ds^2 & = & \rho^2 d \theta^2+ \frac{a^2 \sin^2 \theta}{\rho^2} \left[ dt- \frac{r^2+a^2}{a} d \phi \right]^2+ \frac{\rho^2}{Q(r)}dr^2 \nonumber\\
&& - \frac{Q(r)}{\rho^2}  \left[ dt- {a}\sin^2 \theta d \phi \right]^2, \\
\rho^2 & = & r^2 + a^2 \cos^2 \theta, \\
Q(r) &=& \frac{\kappa F_0}{2}  (1- \beta r^2)^2- 2 m r+r^2+a^2, 
\end{eqnarray}
see \cite{AGarcia2022} for details in the derivation of this solution.

The contravariant metric components that will be used in  the Hamilton-Jacobi equation of the test particle are
\begin{eqnarray}
\label{gtt} 
g^{tt} &= & - \frac{1}{\rho^2 Q(r)} \left[ (r^2 + a^2)^2 -a^2 \sin^2 \theta Q(r) \right], \\
\label{grr} 
g^{rr} &= & \frac{Q(r)}{\rho^2},\\
\label{gtheta} 
g^{\theta\theta} &= & \frac{1}{\rho^2},\\
\label{gphiphi}
g^{\phi\phi} &= & \frac{Q(r) - a^2 \sin^2\theta}{Q(r) \rho^2 \sin^2 \theta},\\
\label{gtphi} 
g^{t\phi} &= &\frac{a}{\rho^2 Q(r)} \left[ Q(r) - (r^2+a^2) \right], 
\end{eqnarray}

\subsection{Horizons}

The horizons are given by the roots  of the polynomial $Q(r)=0$, i.e. by the real positive solutions of

\begin{equation}
\frac{\kappa F_{0}\beta^2}{2}r^{4}+(1-\kappa F_{0}\beta)r^2-2mr+\frac{\kappa F_{0}}{2}+a^2=0.
\end{equation}

\begin{figure}
\includegraphics[scale=0.6]{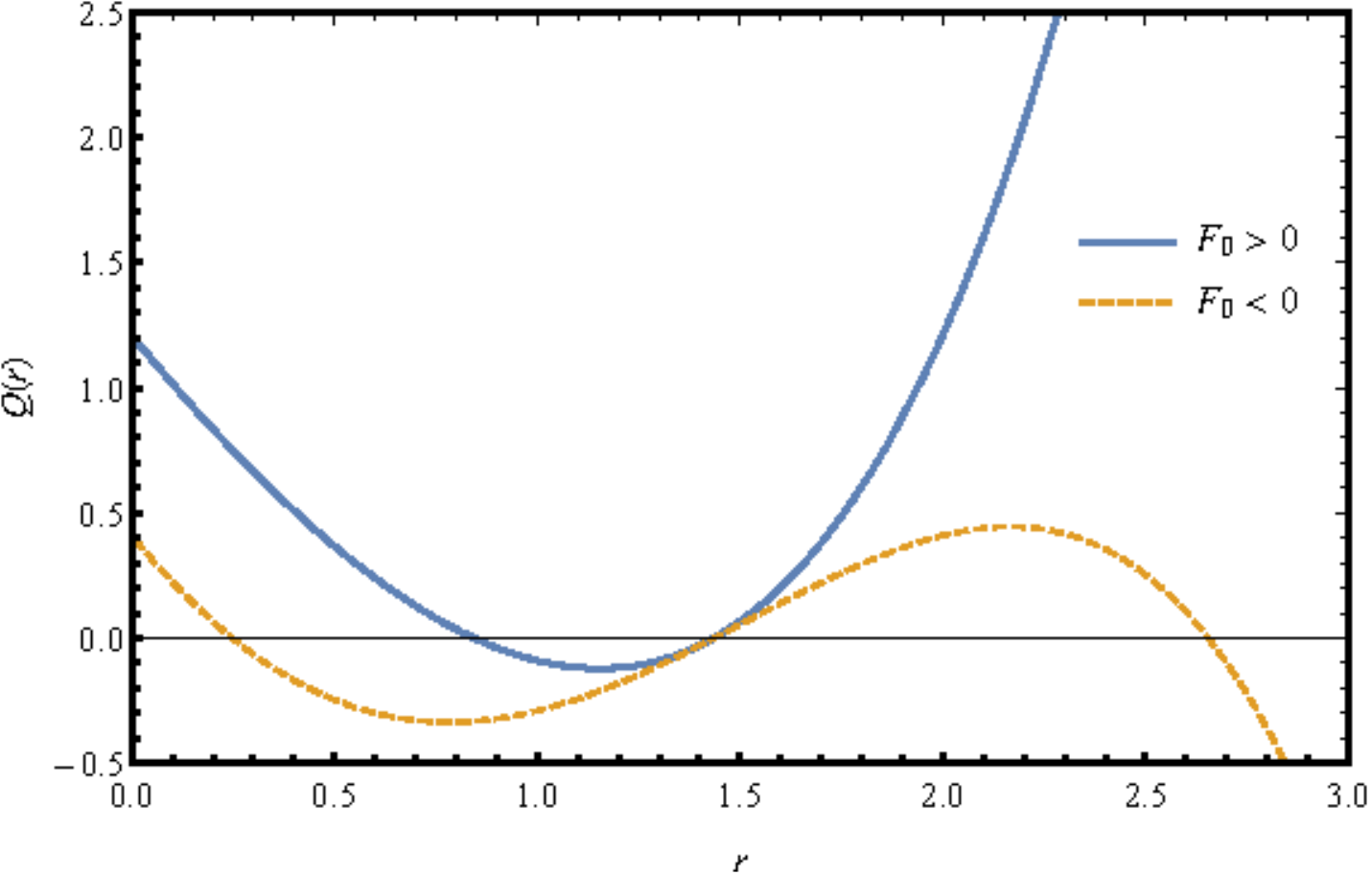}
\caption{The two possible asymptotics of the Stationary  NLE BH are shown; depending on the sign of  $F_0$ the asymptotics of the solution  is different. For $F_0 >0$ there are inner and outer (event) horizons and the asymptotics is AdS; while for $F_0 <0$ the solution presents three horizons: inner, outer and cosmological, 
and the asymptotics is de Sitter. In these plots the fixed parameters are $m=1; \beta=0.5; a=0.9$ and $F_0 = \pm 0.8$.}
\label{Figura1}
\end{figure}

In general there may be four roots  \cite{footnote} depending on the value of $\beta$ and  the sign of the parameter $F_{0}$, that also defines the asymptotics of the solution (see Fig.  \ref{Figura1}). We analyze the two cases of $F_{0}$ being positive or negative.
\begin{itemize}
  \item[(i)] Case $F_{0}>0$. In this case the asymptotic behavior is anti-de Sitter, and the electromagnetic field mimics a  cosmological constant given by
\begin{equation}
\Lambda=-\frac{3}{2}\kappa F_{0}\beta^2.
\end{equation}

There are two values of $r$, $r_{1,2}^{crit}$ and two of $\beta$, $\beta_{1,2}^{crit}$, for which there is only one real root of $Q(r)$. These values are determined from $Q(r^{crit},\beta^{crit})=0$ and $\partial_{r}Q(r^{crit},\beta^{crit})=0$:

\begin{eqnarray}
&& \frac{\kappa F_{0}\beta^2}{2}r^{4}+(1-\kappa F_{0}\beta)r^2-2mr+\frac{\kappa F_{0}}{2}+a^2  = 0,\nonumber\\
&& 2\kappa F_{0}\beta^2r^3+2(1-\kappa F_{0}\beta)r-2m =0.
\end{eqnarray}

From the second expression the values of $\beta$ can be obtained 

\begin{equation}
\beta_{1,2}=\frac{\kappa F_{0}\pm\sqrt{\kappa F_{0}\left[4\left(m-r\right)r+\kappa F_{0}\right]}}{2\kappa r^2F_{0}},
\end{equation}

substituting these values into  the first expression, we arrive at a quadratic equation for $r$ which can be solved using the Cardano-Ferrari method. For the different values obtained from $\beta$, the following cases occur: 

      \begin{itemize}
      \item $\beta<\beta_{1}^{crit}$. There are no real roots. 
      \item $\beta=\beta_{1}^{crit}$. A degenerate real positive root.
      \item $\beta_{1}^{crit}<\beta<\beta_{2}^{crit}$. Two real positive roots.
      \item $\beta=\beta_{2}^{crit}$. A degenerate real positive root.
      \item $\beta>\beta_{2}^{crit}$. There are no real roots.
      \end{itemize}
      
Examples of all these cases are illustrated in Fig.\ref{Figura2}. The corresponding Carter-Penrose diagram of the black hole spacetime (two horizons) is shown in Fig.\ref{Figura3}.
\begin{figure}
\includegraphics[scale=0.6]{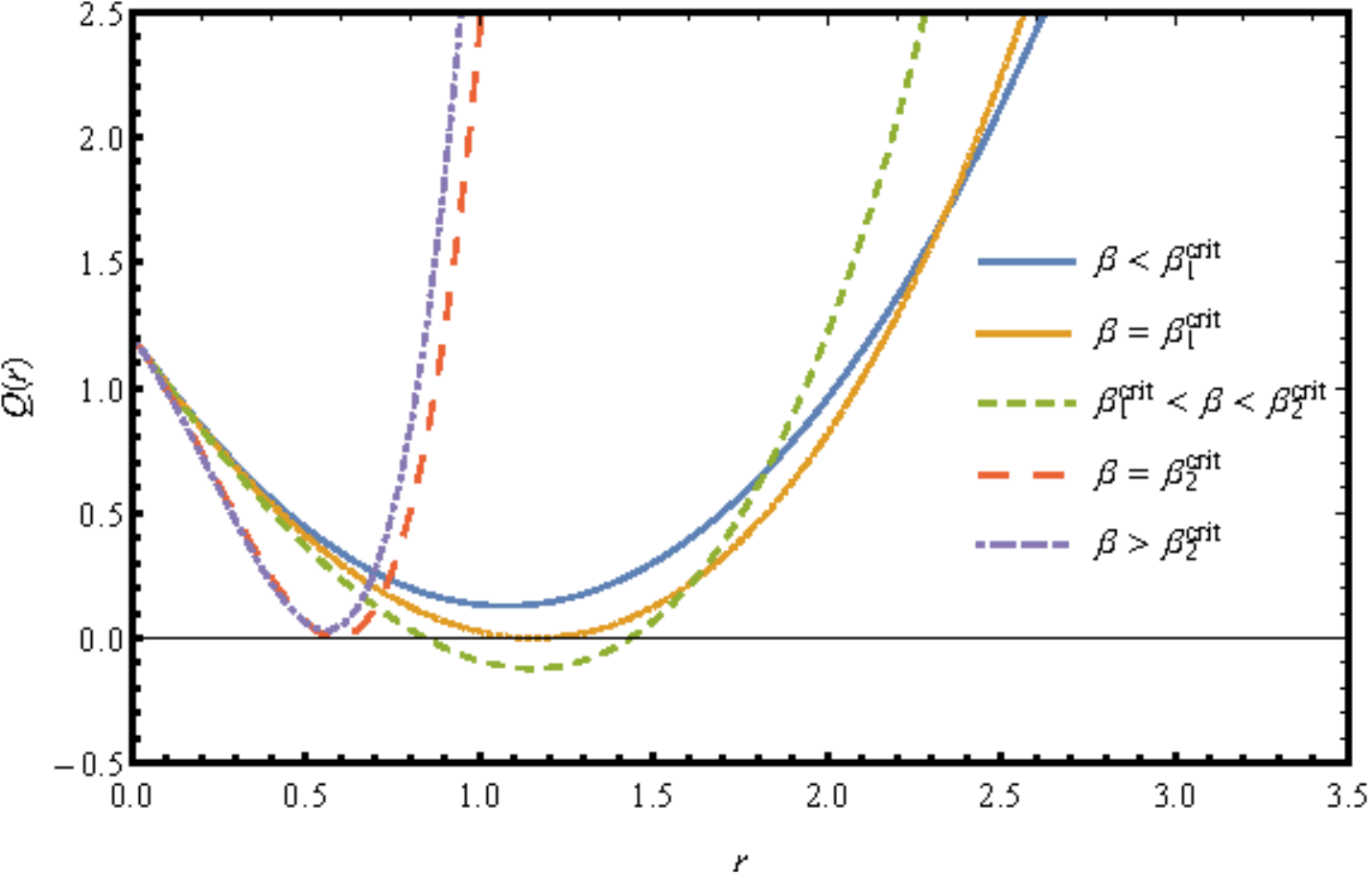}
\caption{The horizons for different values of $\beta$ are shown. For $\beta<\beta_{1}^{crit}$($\beta>\beta_{2}^{crit}$) there are no horizons, for $\beta=\beta_{1}^{crit}$($\beta=\beta_{2}^{crit}$) there is a single degenerate double root, which corresponds to the extreme case with one  horizon at $r_{1}^{crit}\approx 1.15$($r_{2}^{crit}\approx 0.59$). For $\beta_{1}^{crit}<\beta<\beta_{2}^{crit}$ there are two horizons. In this plot we fixed the parameters as $m=1$, $a=0.9$, $F_{0}=0.8$, $\beta_{1}^{crit}\approx 0.26$, and  $\beta_{2}^{crit}\approx 3.54$.}
\label{Figura2}
\end{figure}
  \item[(ii)] Case $F_{0}<0$. In this case the asymptotic behavior is de Sitter, the electromagnetic field  acting as  a cosmological constant given by
\begin{equation}
\Lambda=-\frac{3}{2}\kappa F_{0}\beta^2.
\end{equation}
There are three positive real roots that represent an inner horizon, an outer horizon (event horizon), and a cosmological horizon, c.f. Fig.\ref{Figura1}.
\end{itemize}
In Figs. \ref{Figura1} and \ref{Figura2} the generic behavior of $Q(r)$ is shown. Note that at  the origin, $Q(r=0)$, the metric function is finite, however, and this cannot be illustrated in a 1-dimensional plot, the ring singularity, characteristic of the KN metric, persists in the stationary NLE-BH and occurs when simultaneously $r=0$ and $\theta= \pi/2$, i.e. when $\rho^2= r^2 + a^2 \cos^2 \theta=0$.

\subsection{The ring singularity}

In \cite{Fayos2017}, it is analyzed the rotating Kerr-like spacetime and the conditions for the regularity of their second order polynomial invariants  in the Riemann tensor.  It is shown that the regularity of a Petrov type D spacetime coupled to a non-null electromagnetic field is determined by the finiteness of three invariant function, namely the eigenvalue of the Weyl conformal tensor, the eigenvalue of the traceless Ricci tensor and the curvature scalar, $[\Psi_2, S, R]$
\cite{Gustavo2020}. These invariants, for the Petrov type D solutions,  are related to the Kretschmann quadratic Riemannian invariant by

\begin{equation}
\label{Krets}
\mathcal{K}= R_{\mu \nu \rho \sigma}R^{\mu \nu \rho \sigma}= 48 \Psi_2 \Psi_2^{*
}+8 S^2+ \frac{1}{6}R^2.
\end{equation}

For the metric  (\ref{K_like}) these invariants  $[\Psi_2, S, R]$ are given by

\begin{eqnarray}
&&12 \rho^6 \Psi_2= (a \cos \theta -ir)^2  \left \{ -12 m (r + i a \cos \theta) \right. \nonumber\\
&& +6 \kappa F_0 (1- \beta^2 r^2 a^2 \cos^2 \theta) \nonumber\\
&& \left. + \kappa F_0 \beta [2 ( a^2 \cos^2 \theta- r^2)-8ir a \cos \theta ] \right\},  \nonumber\\
&& 2 \rho^4 S = \kappa F_0 \left( 1+ \beta [-  a^2 \cos^2 \theta (1-3 \beta r^2)+r^2 ] \right), \nonumber\\
&& \rho^2 R= 2 \kappa F_0  \beta  (1-3 \beta r^2), 
\end{eqnarray}
substituting in  Eq. (\ref{Krets}) we obtain  explicitly the  Kretschmann  invariant,

\begin{eqnarray}
&& \mathcal{K}= 48 \frac{m^2}{\rho^6} - 48 \frac{Q_e^2 mr}{\rho^8} +14 \frac{Q_e^4}{\rho^8 } \nonumber\\
&& + 16 \frac{ Q_e^2 \beta mr}{\rho^8}[- 3x ^2(1- \beta r^2)+r^2] \nonumber\\
&& +  \frac{4 Q_e^4 \beta}{\rho^8} \left[ (x^2-r^2)+ \beta (r^4+x^4 + r^2x^2) \right] \nonumber\\
&& + \frac{2 \beta^3 Q_e^4 r^2}{\rho^8} \left[ 3 \beta  r^2 (r^4+2r^2x^2+6x^4) \right. \nonumber\\
&& \left. -  2 (r^4-3r^2x^2+6 x^4)  \right],
\end{eqnarray}
where we have taken $\kappa F_0=Q_e^2$ as the BH electric charge and $x=a \cos \theta$.
From the expression for the    Kretschmann  invariant we see how the introduction of the NLE field affects the curvature. Assuming $\beta >0$ then the curvature is larger than the one of KN, that corresponds to the first three terms. Accordingly, we shall see that for the rotating NLE BH the bounded orbits are closer to the horizon  than for KN.

Moreover, the divergence of the  Kretschmann  invariant points to a real physical singularity at $\rho=0$. We see from the previous expression that the divergence of the Stationary NLE BH  is of the same order than in KN, $\mathcal{K} \approx \rho^{-8}$ and  the inclusion of the NLE field into the Kerr-like metric does not introduce any singularity,
apart of the characteristic ring singularity of the Kerr family, at $\rho=0,$   where the curvature invariant
$ \mathcal{K}$ diverges.  Since, $\rho^2= a^2 \cos^2 \theta + r^2 $ 
vanishes when simultaneously $r=0$ and $\theta = \pi/2$, and recalling  that
$r=0$ is not a point in space but a disc of radius $a^2$, such that the set of points of the singularity is actually the ring at edge of $r=0$. 
This can be seen clearer in ellipsoidal coordinates, see \cite{HawkingEllis} or  \cite{Carroll} for details.

\subsection{Maximal Extension}

The maximal extension for the spacetime (\ref{K_like}) in the case when there are two horizons ($\beta_{1}^{crit}<\beta<\beta_{2}^{crit}$) can be obtained in a similar way to the case of the solution of Kerr-(Newman)\cite{Carter68, HawkingEllis}, we perform a transformation to Kerr coordinates $(r,\theta,u_{\pm},\psi_{\pm})$, where 

\begin{equation}
du_{\pm}=dt\pm\frac{r^2+a^2}{Q(r)}dr,\qquad d\psi_{\pm}=d\phi\pm\frac{a}{Q(r)}dr. 
\end{equation}

The metric  takes the form 

\begin{eqnarray}
&& ds^2=\frac{\left[\left(r^2+a^2\right)^2-Qa^2\sin^2{\theta}\right]}{\rho^2}\sin^2{\theta}d\psi_{\pm}^2 \nonumber\\
&&-\frac{\left(Q-a^2\sin^2{\theta}\right)}{\rho^2}du_{\pm}^2 \nonumber\\
&& -\frac{2a\left(a^2+r^2-Q\right)}{\rho^2}\sin^2{\theta}
du_{\pm}d\psi_{\pm} \nonumber\\
&&+\rho^2 d\theta^2\pm 2 du_{\pm}dr\mp 2a\sin^2{\theta}drd\psi_{\pm}\nonumber\\.
\end{eqnarray}

\begin{figure}
\includegraphics[scale=0.8]{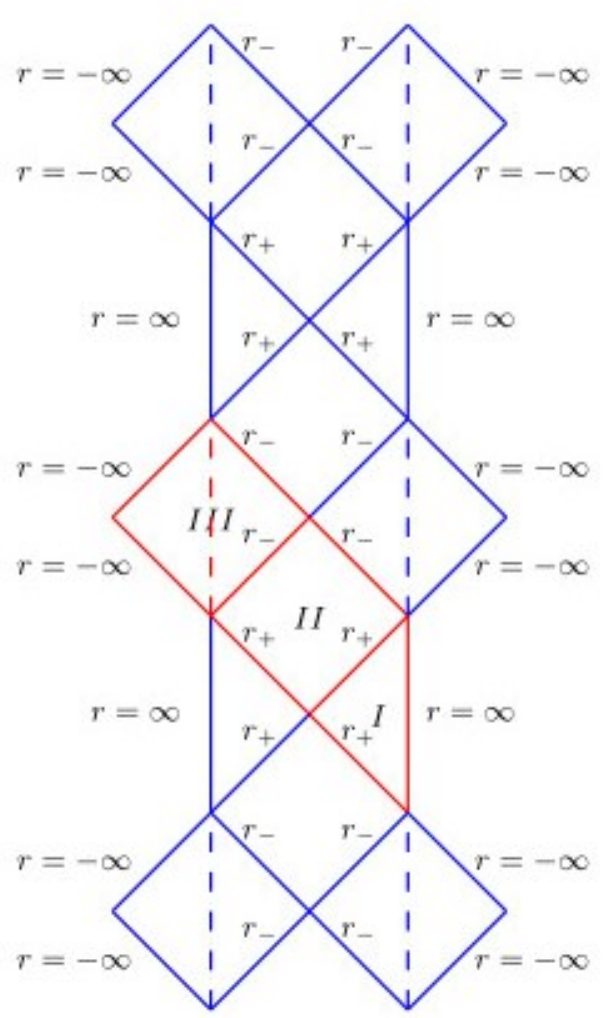}
\caption{The Penrose-Carter diagram of the rotating black hole with nonlinear electrodynamics. The dashed lines represent the ring singularity similar to the Kerr-Newman BH.}
\label{Figura3}
\end{figure}

The maximal analytic extension is built up by a combination of the previous extensions as in the Kerr-Newman case.  The global structure will be very similar to Kerr-Newman. Figure \ref{Figura3} shows the conformal structure of the solution (\ref{K_like}).  Three regions are observed, the regions $I$  represent the regions asymptotically $AdS$ which is $r>r_{+}$.  Regions $II$ ($r_{-}<r<r_{+}$) contain the trapped closed surfaces and finally regions $III$ contain the ring singularity. 

\subsection{Electromagnetic fields}

The  electromagnetic potential  $A^{\mu}$ can be written as a linear combination of two terms, $A_1(\theta)$ depending only on $\theta$ and  $A_2(r)$ depending only on $r$,

\begin{eqnarray}
\rho^2 A_{t} &= &  f_1 a \cos \theta [1+ \beta a^2  \cos^2 \theta ] + g_1 r (1 - \beta r^2) \nonumber\\
&=& A_1 (\theta)+ A_2(r), \nonumber\\
\rho^2 A_{\phi} &= & -  \frac{a^2+r^2}{a}  A_1 (\theta) -  a \sin^2 \theta A_2(r),
\label{empotential}
\end{eqnarray}
and $A_{\theta}=0=A_{r}$;   the electromagnetic fields can be derived from them, being the nonvanishing components $F_{\theta t}, \quad F_{r t}, \quad F_{\theta \phi}, \quad F_{r \phi}$.
Asymptotically the electromagnetic fields  behave as  anti-de Sitter  if $F_0 >0$ or as de Sitter if $F_0 <0$. However if we assume physically reasonable energy conditions there are the constraints that $\beta >0$ and $F_0 >0$  [see \cite{AGarcia2022} for details on energy conditions of this solution].


\section{Motion of charged test particles  in the  stationary NLE BH }

Let us consider a  charged test particle   in the vicinity of the Stationary NLE-BH; the test particle  characterized by a 4-velocity  $u^{\mu}= d x^{\mu}/d \tau$, mass $\mu$ and charge $q$, with $\tau$ being an affine parameter.
Since the spacetime is axisymmetric and stationary, associated with the existence of two Killing vectors,  there are two conserved quantities,
the energy $E$ and the angular momentum  $L_{z}$, related to the 4-velocity as

\begin{equation}
-E=\mu u_{t}+q A_{t}, \quad L_{z}= \mu u_{\phi}+ q A_{\phi}.
\end{equation}

It is then straightforward to determine 
$u^{t}= dt/d \tau= \dot{t}$ and $u^{\phi}=d \phi /d\tau= \dot{\phi}$; 
we define the functions $J(\theta)$ and $T(r),$  that involve the test particle parameters, as

\begin{eqnarray}
J(\theta) &=& a \tilde{L}_{z} - a^2 \sin^2{\theta} \tilde{E}+ \tilde{q} A_{1}(\theta), \nonumber\\
T(r) &=& - a \tilde{L}_{z} +  (a^2 +r^2) \tilde{E}+ \tilde{q} A_{2}(r),
\end{eqnarray}
where the tilde denotes the parameter per unit mass of the test particle, $\tilde{x}= x/ \mu$.
In terms of $J(\theta)$ and $T(r),$    $dt/d \tau= \dot{t}$ and $d \phi /d\tau= \dot{\phi}$ can be written as
\begin{eqnarray}
\rho^2 \dot{t} &=&J(\theta)+ \frac{r^2+a^2}{Q(r)} T(r) , \nonumber\\
\rho^2 \dot{\phi} &=& \frac{J(\theta)}{a \sin^2{\theta}}+ \frac{aT(r)}{Q(r)}.
\end{eqnarray}

Using the Hamilton-Jacobi method we can determine  the remaining components of the test particle 4-velocity, $\dot{r}$ and $\dot{\theta}$, in the  stationary NLE BH.
It turns out that the Hamilton-Jacobi equations are separable in  the $"r"$ and $"\theta"$ coordinates; this fact making manifest that a fourth conserved quantity $K$ exists in the  Kerr-like spacetimes \cite{Visinescu2012} that is the projection of the Killing tensor (also known as the Stackel-Killing tensor), 
$K=K_{\mu \nu} \dot{x}^{\mu} \dot{x}^{\nu}$.

The action is written as:
\begin{equation}
\label{action} S=S_t+S_r+S_\theta+S_\phi - \frac{1}{2} \mu^2 \tau.
\end{equation}
 Based on the mentioned symmetries, some components of the action can be written as:
\begin{equation}
\label{action2} S_t=-Et ;\enspace S_\phi=L_z \phi,
\end{equation}
then  Hamilton-Jacobi equation~is,
\begin{equation}
\label{HJ}
 g^{\mu\nu} p_{\mu} p_{\nu} = g^{\mu\nu}[ \partial_\mu S -qA_\mu][\partial_\nu S-qA_\nu ]= - \mu^2.
\end{equation}
where $q$ is the charge of the test particle and $A_{\mu}$ are the electromagnetic potential components \eqref{empotential}.
Extending the sum in metric components, results in
\begin{eqnarray}
&& g^{tt}(\partial_t S_t  -qA_t)^2+g^{rr}(\partial_r S_r)^2+g^{\theta\theta}
(\partial_\theta S_\theta)^2 \nonumber\\
&& +g^{\phi\phi}(\partial_\phi S_\phi-qA_\phi)^2 \\
&& +2g^{t\phi}(\partial_t S_t-qA_t)(\partial_\phi S_\phi -qA_\phi) =- \mu^2,
\end{eqnarray}

Substituting \eqref{action2}, we have the equation:
\begin{eqnarray}
\label{asterisco}
&&g^{tt}(-E -qA_t)^2+g^{rr}(\partial_r S_r)^2+g^{\theta\theta}
(\partial_\theta S_\theta)^2 \nonumber\\
&& +g^{\phi\phi}(L_z - qA_\phi)^2+ 
2g^{t\phi}(-E - qA_t)(L_z -qA_\phi) =- \mu^2,\nonumber\\
\end{eqnarray}

Substituting the contravariant components $g^{\mu \nu}$,  \eqref{gtt},
\eqref{grr}, \eqref{gtheta}, \eqref{gphiphi}, \eqref{gtphi} into
\eqref{asterisco}, we have:

\begin{eqnarray}
&& Q(r) (\partial_{r} S_{r})^2 -\frac{1}{Q(r)}T(r)^2 +\mu^2r^2= - K \nonumber\\
&&=  - (\partial_{\theta} S_{\theta})^2 - \frac{1}{a^2 \sin^2 \theta} J(\theta)^2 -\mu^2 a^2 \cos^2 \theta.
\label{Sr_eq}
\end{eqnarray}

By separating variables we end up with two 
equations, one depending on
$\theta$,  the $\theta$ part, and the other one depending only on  $r$,  the $r$ part,  with $K$ being the (Carter) separation constant:

\begin{equation}
\label{thetapart}(\partial_\theta S_\theta)^2= K - \frac{1}{a^2 \sin^2 \theta} J(\theta)^2 -\mu^2 a^2 \cos^2 \theta, 
\end{equation}

\begin{equation}  
\label{rpart}   
Q(r) (\partial_rS_r)^2  =   \frac{1}{Q(r)} T(r)^2  -\mu^2r^2 - K.
\end{equation}

It is known that $\partial_\mu S_\mu= p_\mu+qA_\mu=g_{\mu \nu}p^{\nu}+qA_\mu, \quad p^{\nu}=m \dot{x^{\nu}}$ and since $A_{r}=0=A_{\theta}$,
then  $(\partial_\theta S_\theta)$ defines the motion in $\theta$ by the equation:

\begin{eqnarray}
\label{thetapunto}
\left( \rho^2 \dot{\theta} \right)^2 &=& \Theta (\theta) \nonumber\\
&=&
\tilde{K} -a^2 \cos^2 \theta  \nonumber\\
&&
- \frac{1}{a^2 \sin^2 \theta}  [a \tilde{L}_{z} - a^2 \sin^2{\theta} \tilde{E}+ \tilde{q} A_{1}(\theta)]^2,\nonumber\\
\end{eqnarray}
where  $\tilde{K}={K}/{\mu^2}$.

Eq. \eqref{thetapunto}  is still coupled because the  factor  $\rho^2(r, \theta)$  arises in the left hand side of the equation. To decouple it completely we use the orbital  parameter $\lambda$ related with the proper time $\tau$, also known as Mino time and  defined in \cite{Mino2003}, as 

\begin{equation}
\frac{d \lambda}{d \tau} =  \frac{1}{\rho^2}, \quad \tau= \int_{0}^{\lambda} \rho^2 d \lambda.
\end{equation}

Note that the motion equations can be decoupled without  introducing  the Mino time \cite{Carter68}; but using the Mino time  decouples
the radial and colatitudinal equations of motion  in a simpler way.   We present  the forthcoming analysis  in terms of  functions dependent
on the Mino time. The  $\theta$-motion equation  can be written as

\begin{equation}
\frac{d \theta}{d \lambda} = \sqrt{ \Theta (\theta)}.
\end{equation}

The $r$-part of the movement,  from \eqref{Sr_eq}, is determined by the equation
\begin{equation}
\label{rpunto}
\left( \frac{ d r}{d \lambda} \right)^2  =  T(r)^2 - Q(r) \left( \tilde{K} + r^2 \right) = \mathcal{R}(r).
\end{equation}
 
For the massless particle motion, we must take the limit $\mu =0$ in Eqs. (\ref{thetapart})-(\ref{rpart}). Recall that massless test particle trajectories are not the trajectories followed by light rays, since in NLE the latter are governed by the null geodesics of an effective optical metric \cite{Pleban}, that is determined from the NLE Lagrangian.


\subsection{\textbf{$\theta$}-motion}

Let us now consider the $\theta$-motion.
From Eq. \eqref{thetapunto}, we have a restriction for
$\Theta$: that is $\Theta\geq 0$    which is a general
condition to guarantee motion in $\theta$.


First of all we list some general properties of the $\theta$-motion.  From  Eq. \eqref{thetapunto} note that the $\theta-$motion depends on the electromagnetic potential  $A_1(\theta)=f_1 a \cos \theta ( 1 + \beta a^2 \cos^2 \theta)$ related to the magnetic charge of the BH and it has influence only over  a charged test particle since appears as the term $A_{1} \tilde{q}$; then uncharged particles are not affected by the NLE field in their $\theta-$motion and it occurs qualitatively the same as in the Kerr-Newman geometry. Note also that the  $\theta-$motion does not depend on the metric function $Q(r)$. 
Moreover there are some   symmetries evident from  Eq. \eqref{thetapunto}: 

(i) Since the allowed regions for the $\theta-$motion demand that $\Theta \ge 0$,  the Carter constant $K$ must be positive since the subtracting terms in  Eq. \eqref{thetapunto} are positive; then to have an allowed region requires  a larger $K$  than in the KN case ($\beta=0$).

(ii) A simultaneous change of sign of $L_{z}$ and $E$ is equivalent to a reflection on the equatorial plane:   $\Theta |_{- L_{z}, -E} (\theta) = \Theta |_{L_{z}, E } (\pi - \theta)$.

(iii) The movement in the equatorial plane, $\theta= \pi/2$ and $A_{1}=0$,  is not allowed for arbitrary values of $(E, L_{z}, a)$ but it may occur  if $ \tilde{K}- (a \tilde{E}- \tilde{L_{z}})^2 \ge 0$, i. e. the Carter constant  should  be $K \ge (aE-L_{z})^2$.

(iv) A change of sign in $qA_1$ is equivalent to a reflection over the equatorial plane,  $\Theta |_{-q f_1} (\theta) = \Theta |_{q f_1} (\pi - \theta)$. 

(v) One of the poles  cannot be reached by test particles; which one is unreachable,  $\theta=0$ or $\theta= \pi$,   depends on the sign of the product of the BH magnetic charge and the test particle charge   $f_{1}q$; if $\tilde{q} f_{1} >  0$ then test particle cannot reach $\theta=0$, while if  $\tilde{q} f_{1} < 0$ then $\theta= \pi$ is unreachable. Moreover in any case the test particle angular momentum $L_{z}$ cannot be arbitrary   [see Fig. \ref{fig4}], but it must hold that

\begin{equation}
\left( \tilde{q} A_1 (\theta=0, \pi) + a \tilde{L}_{z}  \right)=0, \nonumber\\
\end{equation}
then $ L_{z} = -q A_{1}(\theta=0, \pi)/a$ or $L_{z}=0=q$.

Apart from the previous generalities, 
to examine the allowed regions for the motion in $\theta$ in terms of the BH and test particle parameters,  we proceed to a change of
variable from $\theta$ to $x$ as follows:
\begin{equation}
\label{thetatoxi} x=   \cos\theta,
\end{equation}
with range $-1 \le x \le 1$;  the new variable $x$ in terms of the old
variable $\theta$:
\begin{equation}
\dot{x}= - \dot{\theta}  \sin\theta, \quad \dot{x}= - \sqrt{\Theta}  \sin\theta,
\end{equation}
 
 \begin{figure}
 \centering
 \includegraphics[scale=0.35]{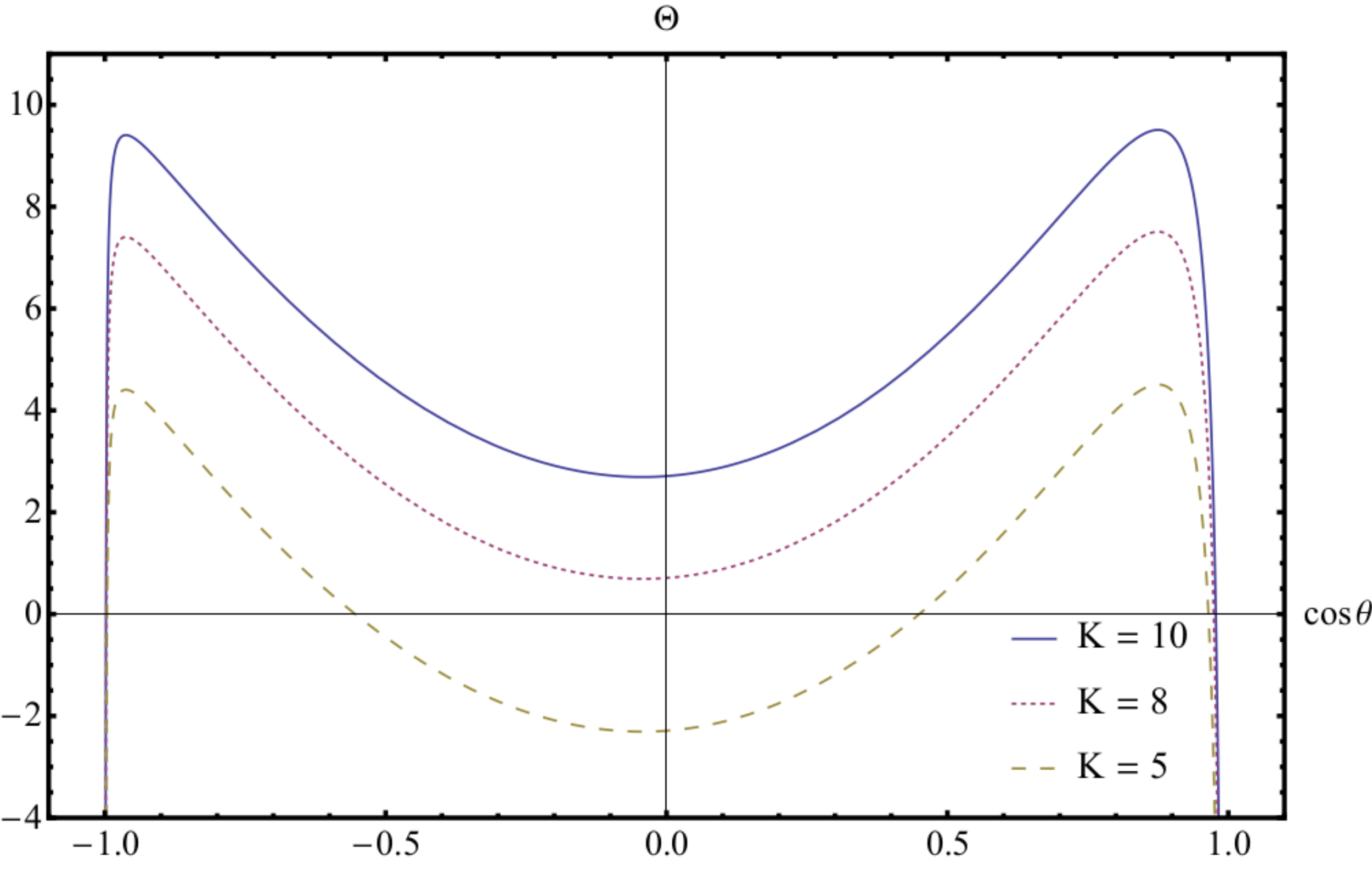}
 \caption{\small The available regions for the bounded orbits in $\theta$-motion are shown  for three values of the Carter constant $K$. The regions $\Theta \ge 0$ are the ones allowed for the motion of test particles and  $K$ must be large enough to  make $\Theta \ge 0$, from top to bottom $K$ decreases as $K=10, 8, 5$. In this plot we fixed the parameters as $m=1; \quad a=0.8; \quad  \beta=5;  \quad f_1=0.5$ (BH magnetic charge); $\tilde{q}=0.3; \quad \tilde{E}=4; \quad \tilde{L}_{z}=0.5.$  Note that for these fixed parameters,  if $\tilde{K} =5$  there are two disconnected regions $\Theta > 0$, one  above and the other below the equatorial plane where the test particle is allowed to move in bounded orbits, and there is a region not available for the test particle motion around the equatorial plane, $-0.5 < \cos \theta < 0.45$. Also note that the regions are not symmetrical respect to the equatorial plane and that the allowed region does not include $\theta= 0$ ($\cos \theta =1$). }
 \label{fig1}
 \end{figure}
 \begin{figure}
 \centering
 \includegraphics[width=0.4\textwidth]{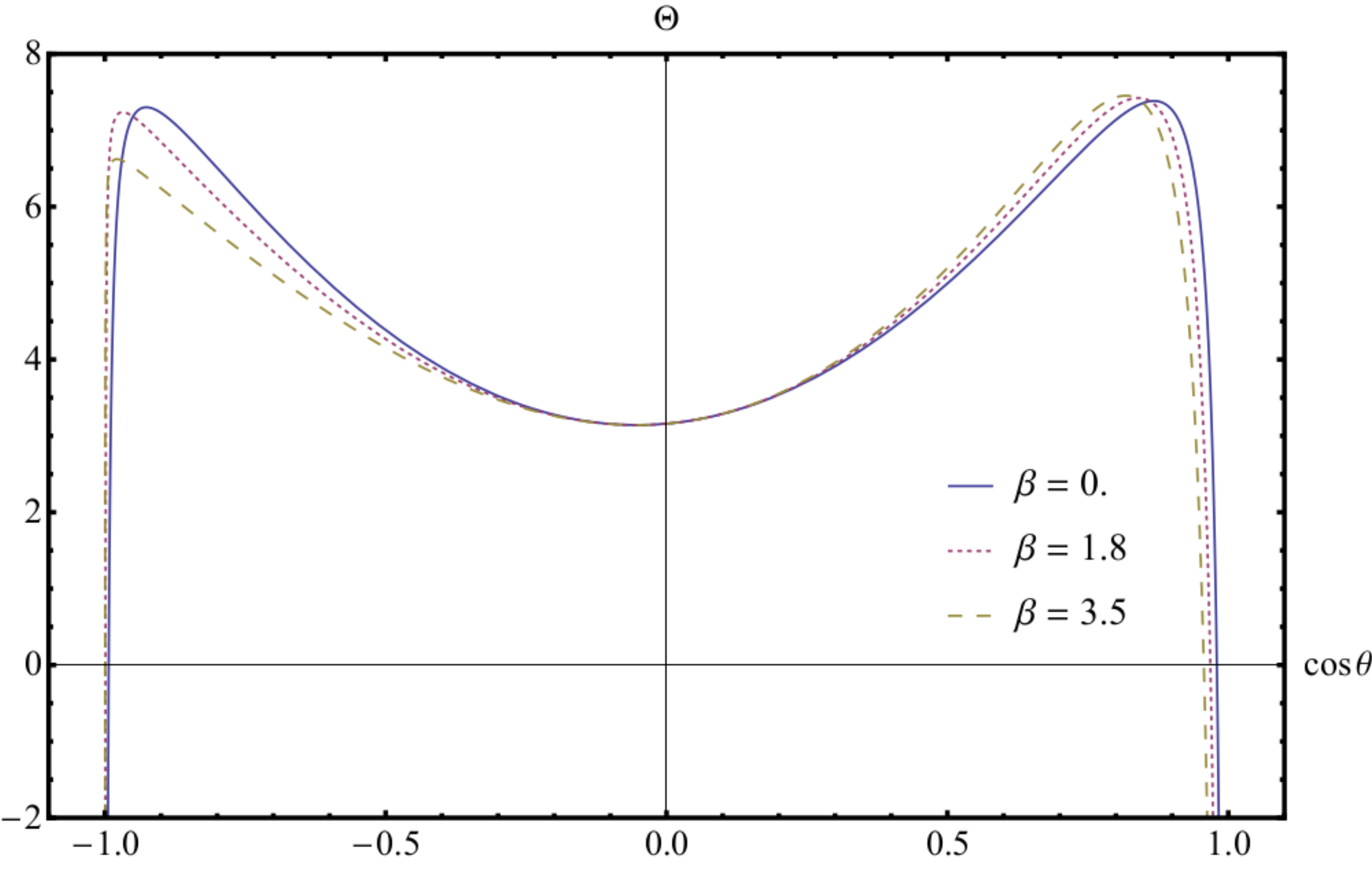}
 \caption{\small   The available regions  for the bounded orbits in  $\theta$-motion are  shown for  three values of the nonlinear parameter $\beta$,  from top to bottom  $\beta= 0, 1.8, 3.5$;   for fixed parameters  the allowed regions   are not very different from the  KN-BH ($\beta=0$); the situation is not different with $\beta$ negative. Note that even for $\beta=0$ the allowed regions are not symmetrical respect to the equatorial plane ($\theta= \pi/2$) and as $\beta$  increases the asymmetry in enhanced. In this plot we fixed the parameters as $m=1; \quad  f_1=0.5$ (BH magnetic charge); $\quad a=0.9; \quad \tilde{q}=0.3$; $\tilde{K}=8; \quad \tilde{E}=3; \quad \tilde{L}_{z}=0.5$. }
 \label{fig2}
 \end{figure}
 \begin{figure}
 \centering
 \includegraphics[width=0.4\textwidth]{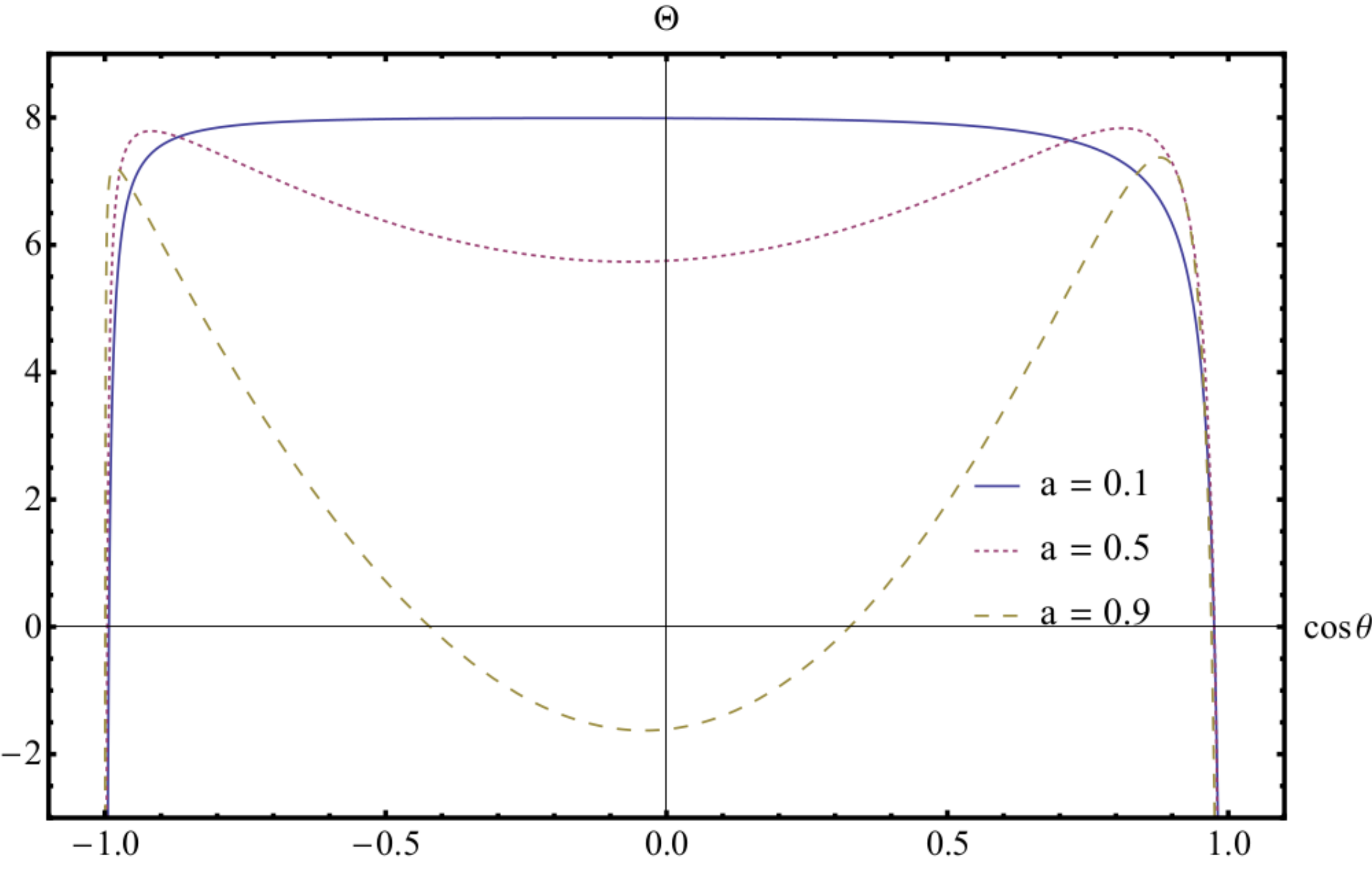}
 \caption{\small   The available regions  for the bounded orbits in  $\theta$-motion are shown for three values of the BH angular momentum, from top to bottom  $a=0.1, 0.5, 0.9$. For small values of $a$ the motion is allowed almost in the whole $\theta$ range, and increasing $a$ restricts the allowed region; see that for $a=0.9$ the allowed region splits into two regions one of them above and the other below the equatorial plane.  In this plot we fixed the parameters as $m=1; \quad  f_1=0.5$ (BH magnetic charge); $\tilde{q}=0.3;$
 $\beta=2; \quad  \tilde{E}=3; \quad \tilde{L}_{z}=0.5; \quad K=8$. }
 \label{fig3}
 \end{figure}
\begin{figure}
 \centering
 \includegraphics[width=0.4\textwidth]{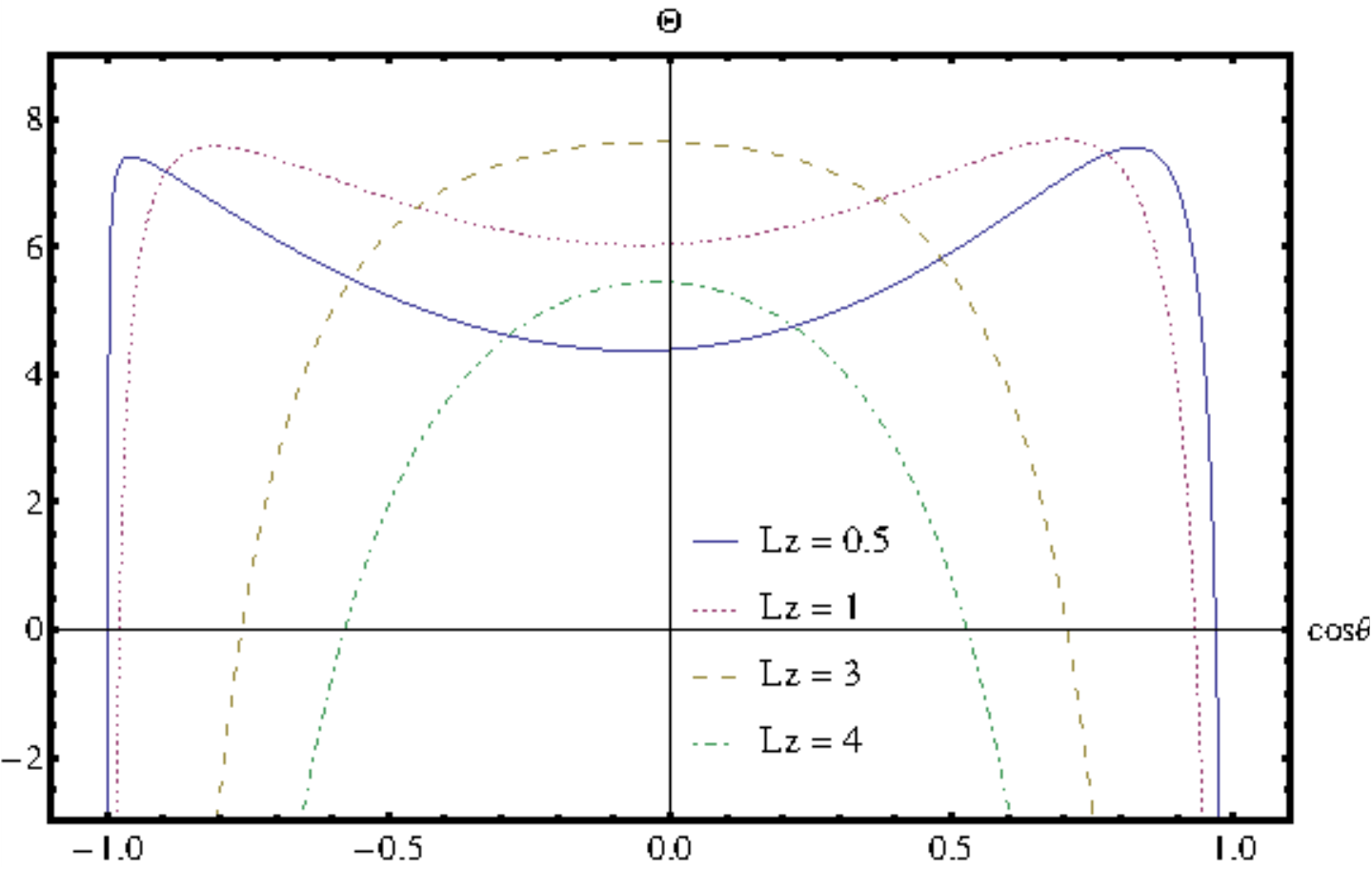}
 \caption{\small   The available regions  for the bounded orbits in  $\theta$-motion, $\Theta \ge 0$,  are shown for different values of the test particle angular momentum $\tilde{L}_{z}.$   The allowed regions decrease their $\theta$ range  as $L_{z}$ increases, for fixed values of the rest of parameters; note that for large values of $L_{z}$ the motion is constrained to a region near the equatorial plane.  In this plot we fixed the parameters as $m=1; \quad a=0.8; \quad f_1=0.5$ (BH magnetic charge); $\beta=2; \quad  \tilde{E}=3; \quad  \tilde{q}=0.3; \quad \tilde{K}=8.$  }
 \label{fig4}
\end{figure}
\begin{figure}
 \centering
 \includegraphics[width=0.4\textwidth]{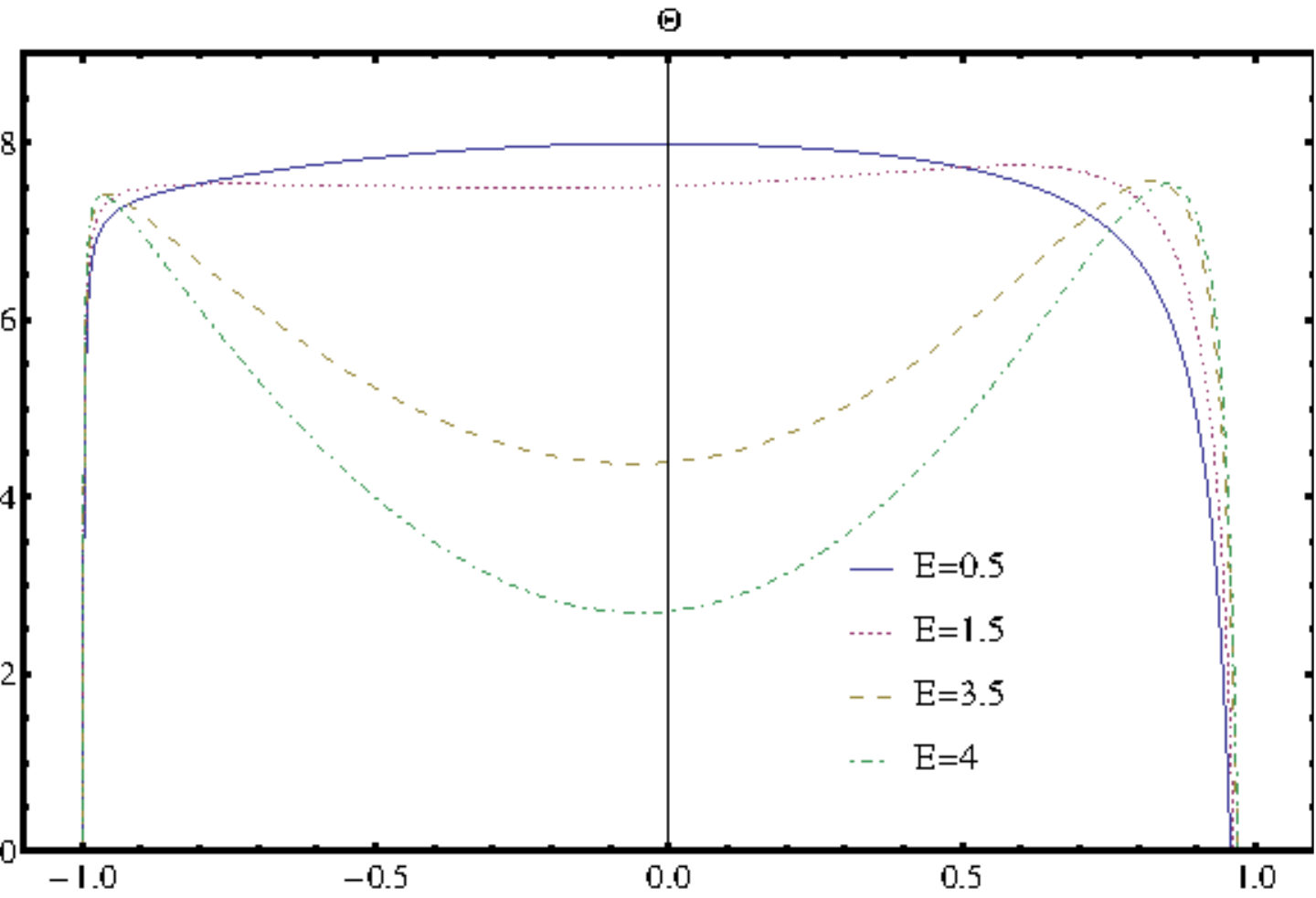}
 \caption{\small   The available regions  for the bounded orbits in  $\theta$-motion are shown for different values of the test particle energy $\tilde{E}.$ 
 The motion is allowed almost in the whole $\theta$ range, with the exception of the pole at $\theta=0$; varying $\tilde{E}$ changes the value of $d \theta /d \lambda$ at the equatorial plane ($\theta= \pi/2$), such that increasing $\tilde{E}$ decreases the angular velocity of the test particle,  $d \theta /d \lambda$  in the equatorial plane and  its vicinity.
 In this plot we fixed the parameters as $m=1; \quad a=0.8; \quad \tilde{q}=0.3; f_1=0.5$ (BH magnetic charge); $\beta=5; \quad \tilde{L}_{z}=0.5; \quad \tilde{K}=8.$  }
 \label{fig5}
\end{figure}

Then the  Eq. (\ref{thetapunto}) in terms of  $x$,  renders a sixth degree polynomial, 
\begin{eqnarray}
\Theta (x)&=& - x^6 a^2 q^2 \beta^2 f_1^2 -  x^5  2 a^3 \beta E f_1 q \nonumber\\
&& + x^4  a^2 (1-E^2-2 \beta q^2 f_1^2) \nonumber\\
&&
+ x^3 2 a qf_1 [- E  + a \beta (a E -  L_{z})] \nonumber\\
 && + x^2 \left[ - K - f_1^2 q^2 - a^2  +2a E(aE-L_{z}) \right] \nonumber\\
&&
 + x 2 qf_1(aE-L_{z}) -a^2 (aE-L_{z})^2 +  K,\nonumber\\ 
\end{eqnarray}
where the tilde $\tilde{}$ has been omitted;
and the condition that $\Theta \ge 0$ defines the  regions where the $\theta-$movement is available. The previous equation can be written as

\begin{eqnarray}
\Theta&=& - x^3  \beta a^2 q f_1 \left[  \beta f_1 q x^3 + 2 a E x^2 +  2 f_1q x - 2 (aE - L_{z}) \right] \nonumber\\
&&
+  x^4  a^2 (1-E^2) - x^3 2 a qf_1 E + \nonumber\\
 && x^2 \left[ -  K - f_1^2 q^2 - a^2  +2a E(aE-L_{z}) \right]\nonumber\\
&&
 + x 2 qf_1(aE-L_{z}) -a^2 (aE-L_{z})^2 +  K, 
\end{eqnarray}
where we have separated the nonlinear contribution in the first term; note that the nonlinear parameter appears combined as $a^2 \beta f_1 q$  and in general we expect this term be small since $a$ is restricted to be less that the BH mass, as well as the magnetic (electric) BH charge; while the reduced test particle charge $\tilde{q}=q/m$  cannot be large, then the NLE  effect  in the $\theta$-motion is not  very significant.   In   Figs 4-8 are shown the allowed regions, $\Theta (\theta)  \ge 0$ when varying the parameters $\tilde{K},  \beta, a, \tilde{L}_{z}$, and $\tilde{E}$.
Recall that  BH parameters are: mass, $m$;  magnetic charge  $f_1$; electric charge, $g_1$;  electromagnetic nonlinear parameter $\beta$. While the test particle parameters are: mass, $\mu$;  electric  charge $q$, energy, $\tilde{E}$;  the angular momentum projection on $z$-axis, $\tilde{L}_{z}$; and Carter constant  $K.$   The latter is associated with the total angular momentum: the total angular momentum of the test particle is not conserved and varies with $\theta$; but the total angular momentum  (BH angular momentum plus test particle angular momentum) must be conserved, so there is an exchange of angular momentum between the BH and the test particle, analogous to the KN case.

From the plots we see that depending on the values of the parameters the bounded movement in $\theta$  occurs in three types of  regions: (i) the first including  the whole range  $0 \le \theta  \le \pi$, excepting one of the poles; (ii) in a strip defined by  $\theta_{\rm min} < \theta < \theta_{\rm max}$; (iii) the allowed region splits into two  disconnected regions one above and the other below the equatorial plane, $\theta_1 < \theta < \theta_2$ and 
$\theta_3 < \theta < \theta_4$ where $\theta_i$ are positive real  roots  of $\Theta(\theta_i)=0$, with general  features described at the beginning of this section;  in all cases one of the poles is unreachable.  The movement in $\theta$ being determined by the BH magnetic charge, while the  BH electric charge  does  not have any influence in this motion.   The Eq. (\ref{thetapunto}) for the   linear (KN) limit obtained with  $\beta=0$  is a fourth order polynomial given by
\begin{eqnarray}
\Theta&=&x^4  (1-E^2) - x^3 2q f_1  E + \nonumber\\
 && x^2 \left[ - K - f_1^2 q^2 - a^2  +2a E(aE-L_{z}) \right] \nonumber\\
&&
 + x 2a qf_1(aE-L_{z}) -a^2 \left(  (aE-L_{z})^2 - K \right),\nonumber\\ 
\end{eqnarray}
that has been thoroughly studied by Hackmann in \cite{Hackmann2013}.
\begin{figure}
 \centering
 \includegraphics[width=0.4\textwidth]{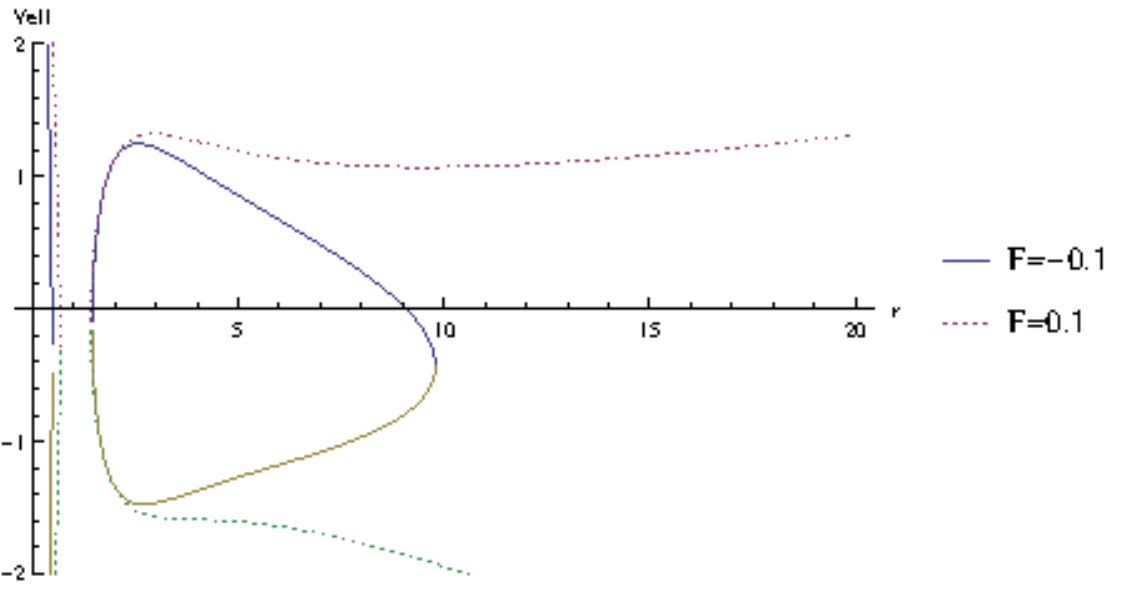}
 \caption{\small The curves show the effective potentials  $V_{\rm eff  \pm}$
 for $F_0= -0.2, 0.2$. The area between the curves is forbidden for the motion of the test particle.
Note that   $V_{\rm eff  -}$ can be negative, then the possibility exists of energy extraction by means of the Penrose process.
In this plot we fixed the parameters as $m=1; \quad a=0.9; \quad \tilde{q}=0.3; g_1=0.5$ (BH magnetic charge); $\beta=0.3; \quad  \tilde{L}_{z}=0.5; \quad \tilde{K}=40$  }
 \label{figEffPot1}
\end{figure}
\begin{figure}
 \centering
 \includegraphics[width=0.4\textwidth]{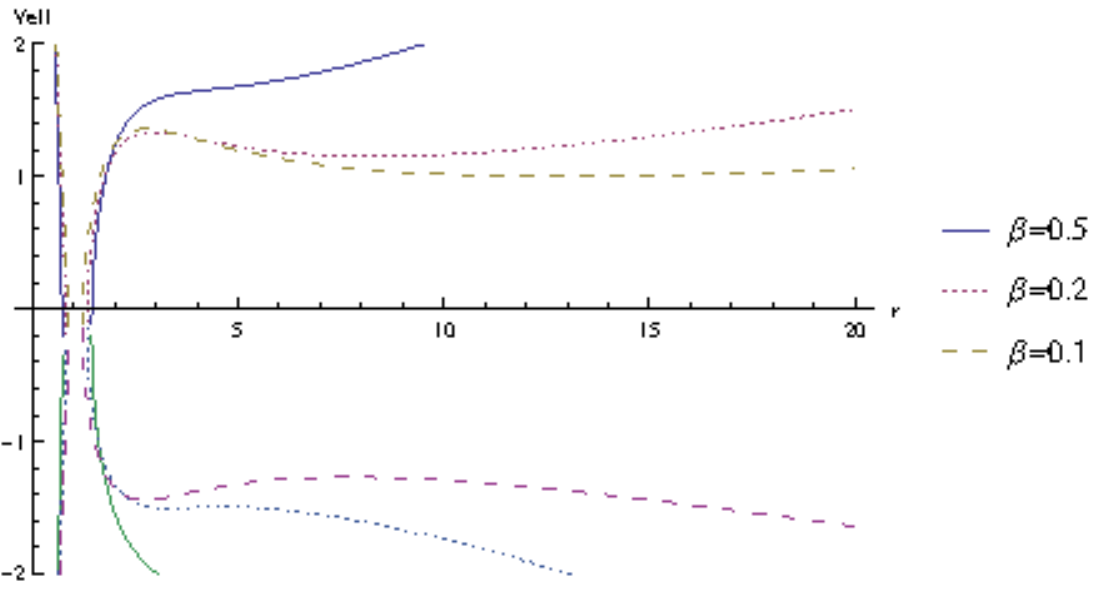}
 \caption{\small The effective potentials $V_{\rm eff  \pm}$ are shown for three different values of  $\beta$. The area between the curves is forbidden for the movement of the test particle.
 In this plot we fixed the parameters as $m=1; \quad a=0.9; \quad \tilde{q}=0.3; g_1=0.5$ (BH magnetic charge); $F_0=0.2; \quad  \tilde{L}_{z}=0.5; \quad \tilde{K}=40$. There exist both maxima and minima then unstable and stable circular orbits occur; as $\beta$ increases no minima occur, then no circular orbits are present for $\beta >0.5$, for the chosen values of parameters in this plot.  }
 \label{figEffPot2}
\end{figure}
\begin{figure}
 \centering
 \includegraphics[width=0.4\textwidth]{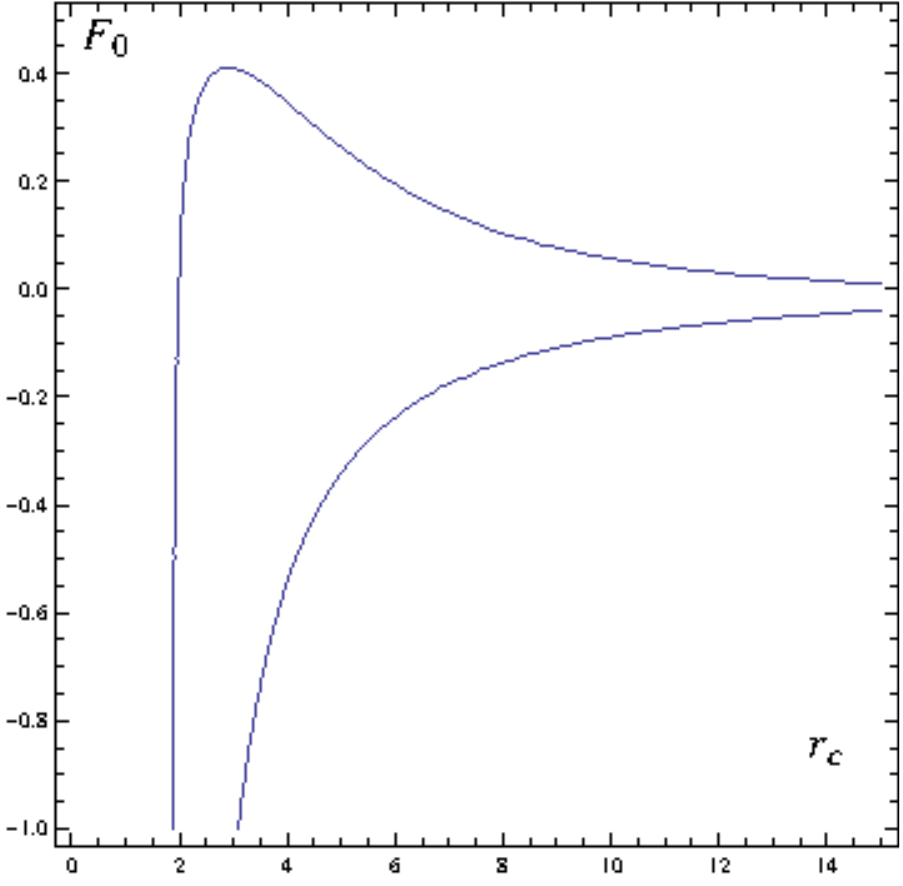}
 \caption{\small The curve shows  $F_0$  (related to the BH charge)  as a function of the radius of the circular orbits  $r_c$.
 In this plot we fixed the parameters as $m=1; \quad a=0.9; \quad \tilde{q}=0.3; g_1=0.5$ (BH electric charge); $\beta=0.3; \quad  \tilde{L}_{z}=0.5; \quad \tilde{K}=40$. There are values of  $F_0  > 0$ that allow two  circular orbits while none exists for $F_{0} > 0.43$. }
 \label{figX}
\end{figure}
\begin{figure}
 \centering
 \includegraphics[width=0.4\textwidth]{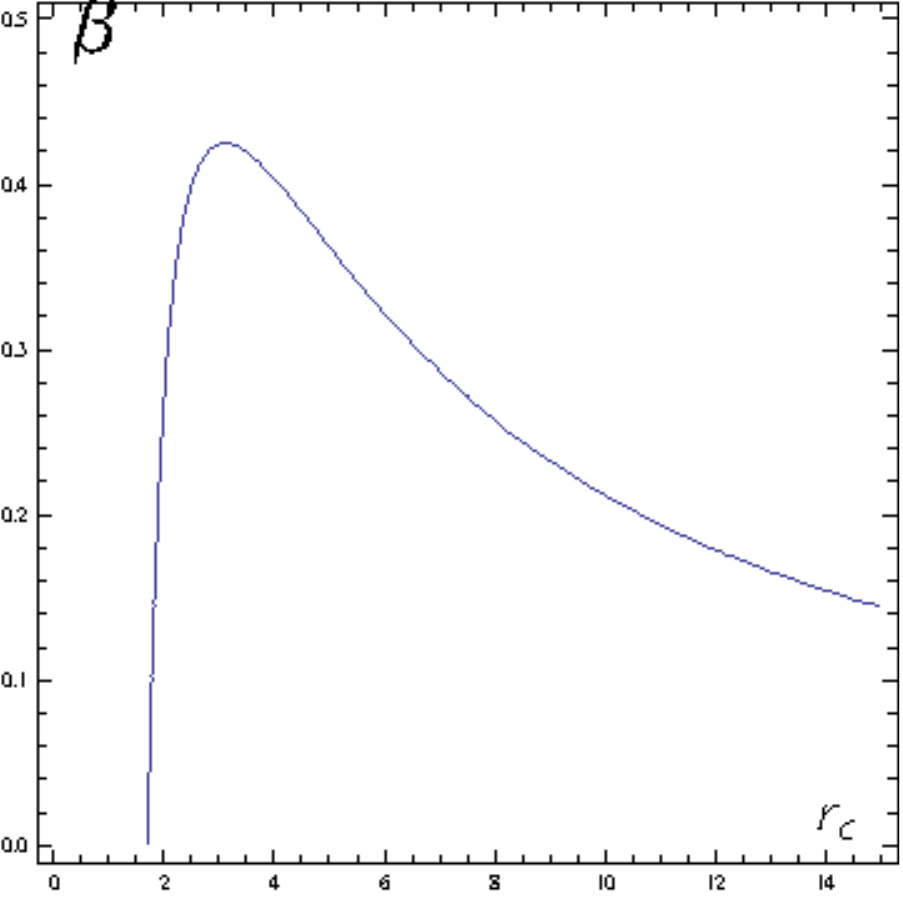}
 \caption{\small The curve shows the NLE parameter  $\beta$ as a function of  the radius of the circular orbits  $r_c$.
 In this plot we fixed the parameters as $m=1; \quad a=0.9; \quad \tilde{q}=0.3; g_1=0.5$ (BH  electric charge); $F_0=0.2; \quad  \tilde{L}_{z}=0.5; \quad \tilde{K}=40$. Note that there  are two circular orbits for  $\beta$  in the range  $0.14  < \beta <  0.43$ and there are not circular orbits   for $\beta> 0.43$. In the case KN, $\beta=0$, there is only one circular orbit.  }
\label{figY}
\end{figure}
\begin{figure}
 \centering
 \includegraphics[width=0.4\textwidth]{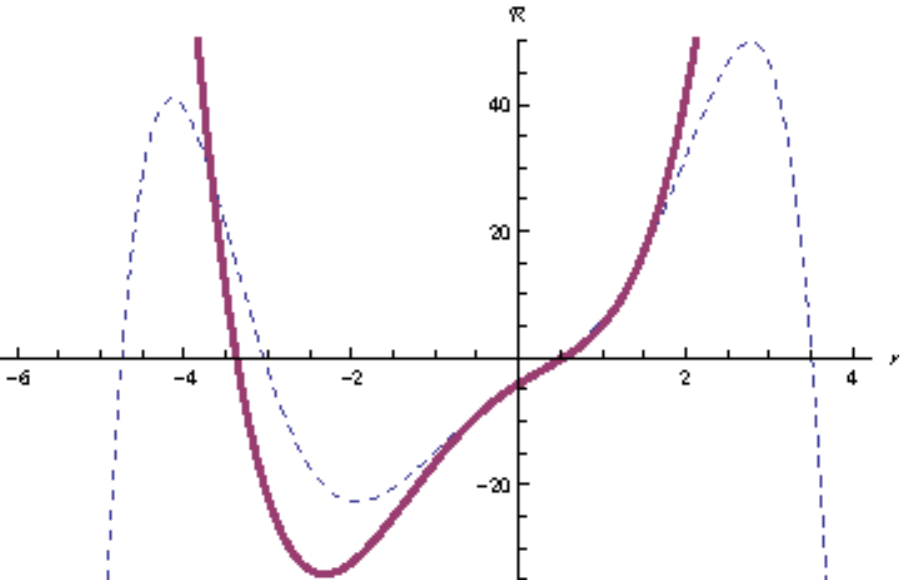}
 \caption{\small The generic behavior of the function  $\mathcal{R}$ is illustrated for the (continuous) Kerr-Newman (KN) ($\beta=0$) and for the (dashed)  Stationary NLE BH ($\beta=0.5$).  For the KN BH $\mathcal{R}=0$ has  two real roots, one positive  and   one  negative; while  the Stationary NLE BH
$\mathcal{R}=0$ has  four real roots, two positive and two negative;
in addition of two complex conjugate roots. The two positive roots are the turning points and the  radius that  delimit the region of bounded orbits. This region is related to the BH capacity of growing an accretion disk. 
 In this plot the parameters are fixed as $m=1; \quad \tilde{L}_{z}=0.5; \quad \tilde{q}=0.3; \quad \tilde{K}=4;  \quad a=0.9; \quad F_0=0.36; \quad  g_1=0.6; \quad  E=1.5$. }
\label{figZ}
\end{figure}
\begin{figure}
\centering
\includegraphics[width=0.4\textwidth]{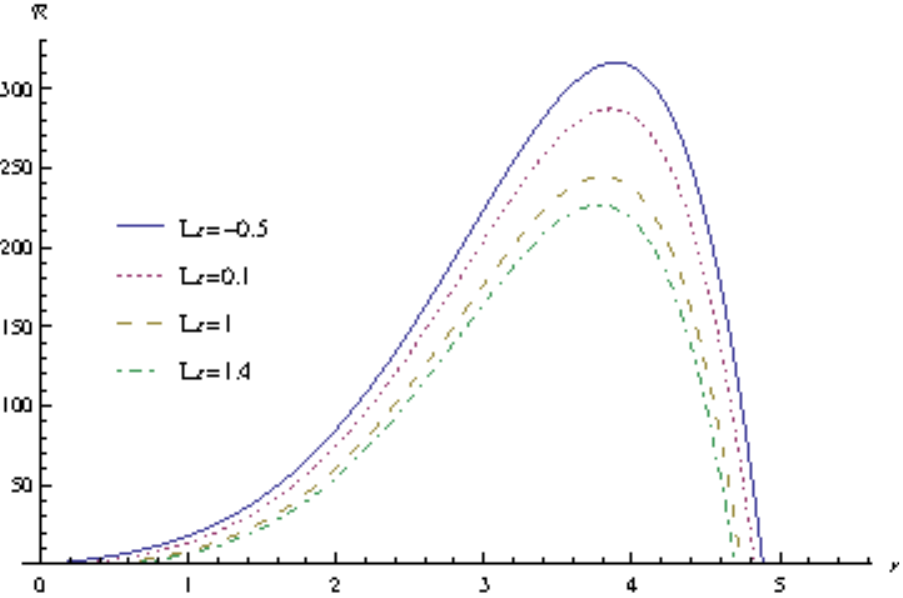}
\caption{\small     The allowed regions for bounded orbits, $\mathcal{R} \ge 0$,   are shown for different values of the test particle angular momentum, from top to bottom $\tilde{L}_{z}= -0.5, 0.1, 1, 4$; the allowed region decreases as $\tilde{L}_{z}$ increases. The rest of the parameters are fixed as $m=1; \quad \tilde{E}=2; \quad \tilde{q}=0.3; \quad \tilde{K}=4; \quad  m=1; \quad a=0.9;  \quad \beta=0.5; \quad F_0=0.36; \quad  g_1=0.6 $
}
\label{figLz}
\end{figure}

\begin{figure}
\centering
\includegraphics[width=0.4\textwidth]{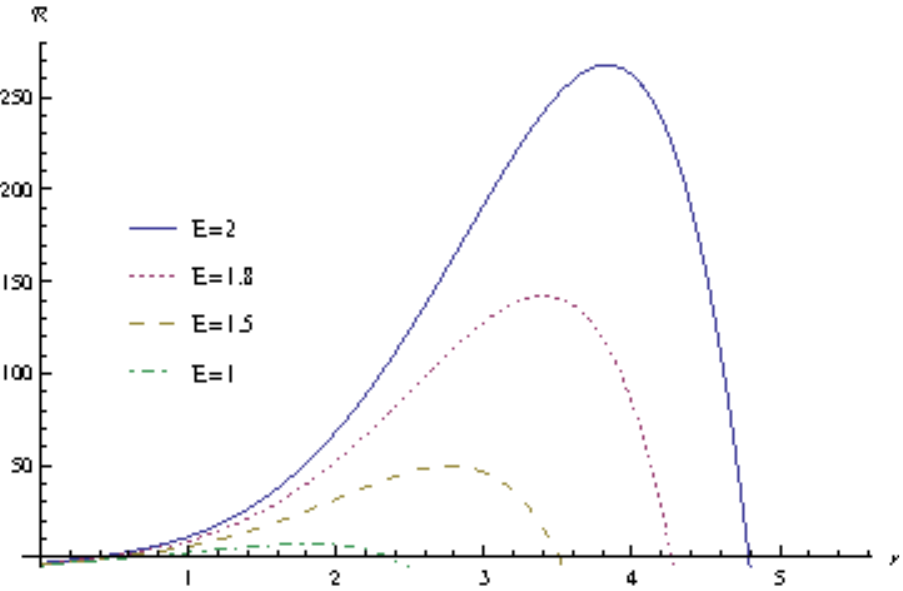}
\caption{\small     It is shown $\mathcal{R}$ as a function of $r$;  the allowed regions  for bounded orbits, $\mathcal{R} \ge 0$, are shown   for different values of the test particle energy, from top to bottom $\tilde{E}= 2, 1.5, 1$; the allowed region increases as $E$ augments; note that for $E<1$ no bounded orbits are possible. The rest of the parameters are fixed as $m=1; \quad \tilde{L}_{z}=0.5; \quad \tilde{q}=0.3; \quad \tilde{K}=4;  \quad a=0.9;  \quad \beta=0.5; \quad F_0=0.36; \quad  g_1=0.6 $
}
\label{figE}
\end{figure}
\begin{figure}
\centering
\includegraphics[width=0.4\textwidth]{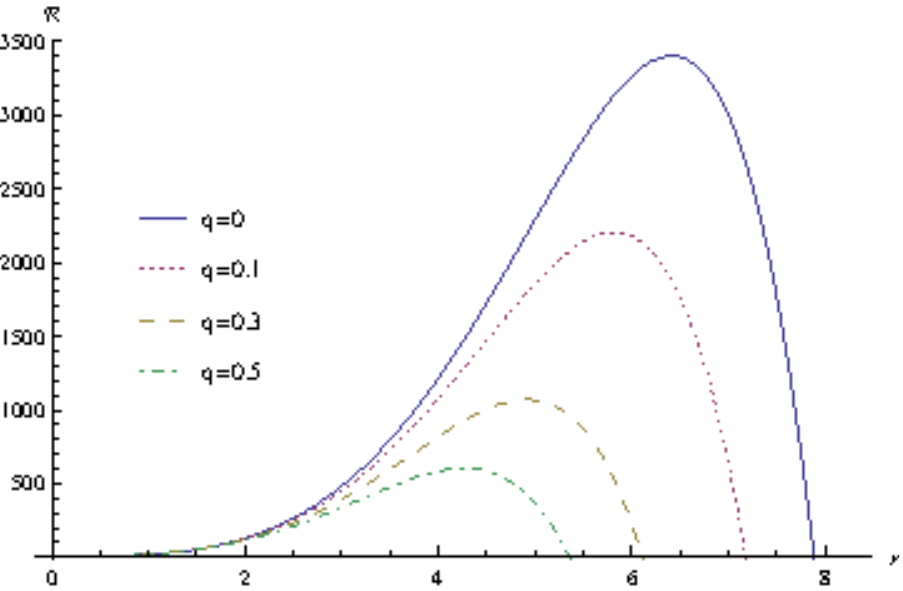}
\caption{\small    The regions allowed for a bounded test particle motion, $\mathcal{R} \ge 0$, are shown   for different values of the test particle charge $\tilde{q}=q/m$. The rest of the parameters are fixed as $m=1; \quad E=2.5; \quad \tilde{L}_{z}=0.3; \quad \tilde{K}=4; \quad  M=1; \quad a=0.9;  \quad \beta=0.5; \quad F_0=0.36; \quad  g_1=0.6$. The uncharged particle has the largest allowed region. 
}
\label{figq}
\end{figure}
\begin{figure}
\centering
\includegraphics[width=0.4\textwidth]{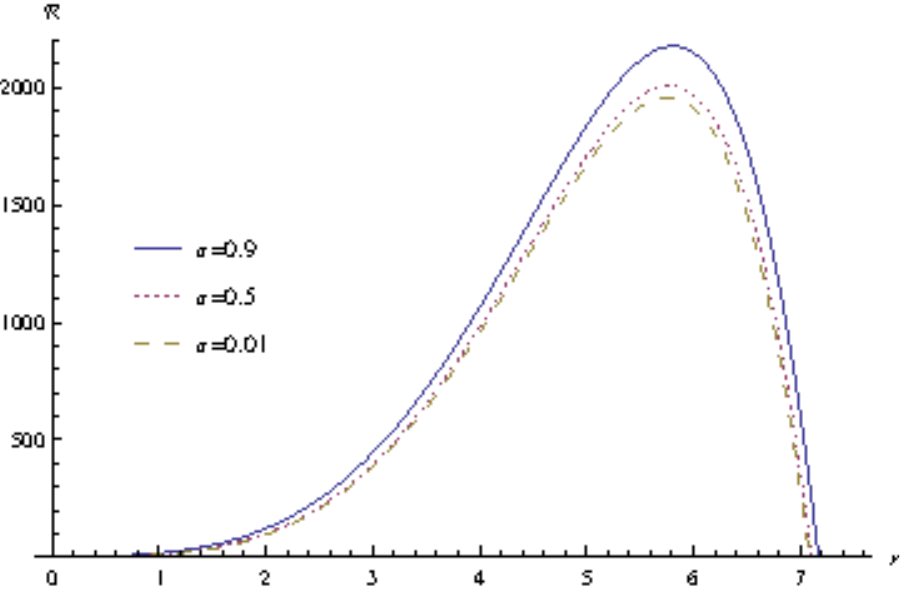}
\caption{\small     The allowed regions  for bounded orbits, $\mathcal{R} \ge 0$,  are shown for different values of the BH angular momentum, from top to bottom $a= 0.9, 0.5, 0.3, 0.1$; as $a$ decreases the allowed region gets smaller but the reduction is not substantial. The rest of the parameters are fixed as $m=1; \quad \tilde{E}=2.5; \quad \tilde{q}=0.1; \quad  \tilde{L}_{z}=0.5;  \quad \beta=0.5; \quad F_0=0.36; \quad  g_1=0.6; $
}
\label{figa}
\end{figure}
\begin{figure}
\centering
\includegraphics[width=0.4\textwidth]{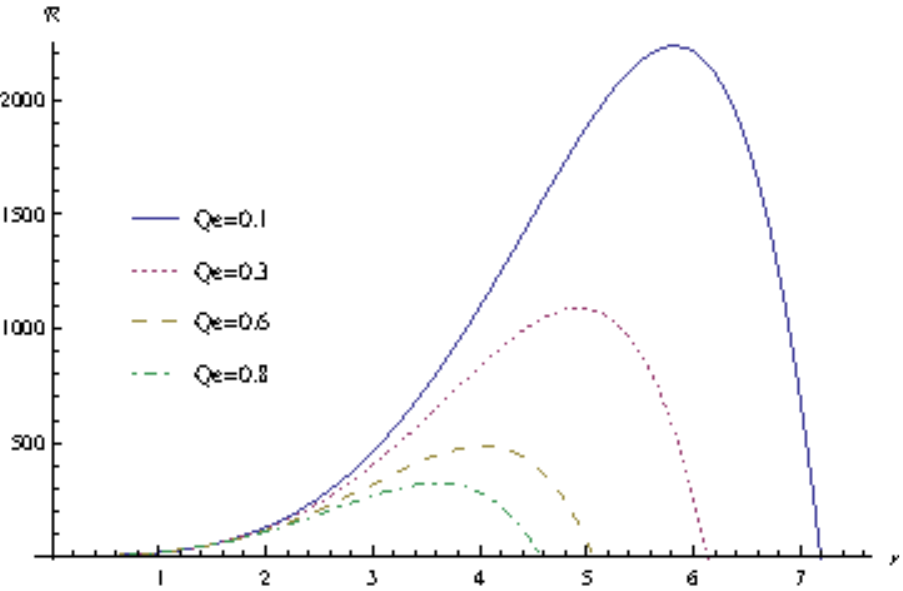}
\caption{\small     It is shown $\mathcal{R} $ as a function of $r$, for different values of the BH electric charge $g_1=Q_e= 0.1, 0.5, 0.6, 0.8$; the allowed regions  for bounded orbits, $\mathcal{R} \ge 0$,  decrease as $Q_e$ increases; for the uncharged BH (not shown in this plot) there are not  test particle bounded orbits.  The rest of the parameters are fixed as $m=1; \quad \tilde{E}=2.5; \quad \tilde{q}=0.1; \quad  \tilde{L}_{z}=0.5;  \quad \tilde{K}=4; \quad a=0.9; \quad \beta=0.5; \quad F_0=0.36 $
}
\label{figQ}
\end{figure}
\begin{figure}
\centering
\includegraphics[width=0.4\textwidth]{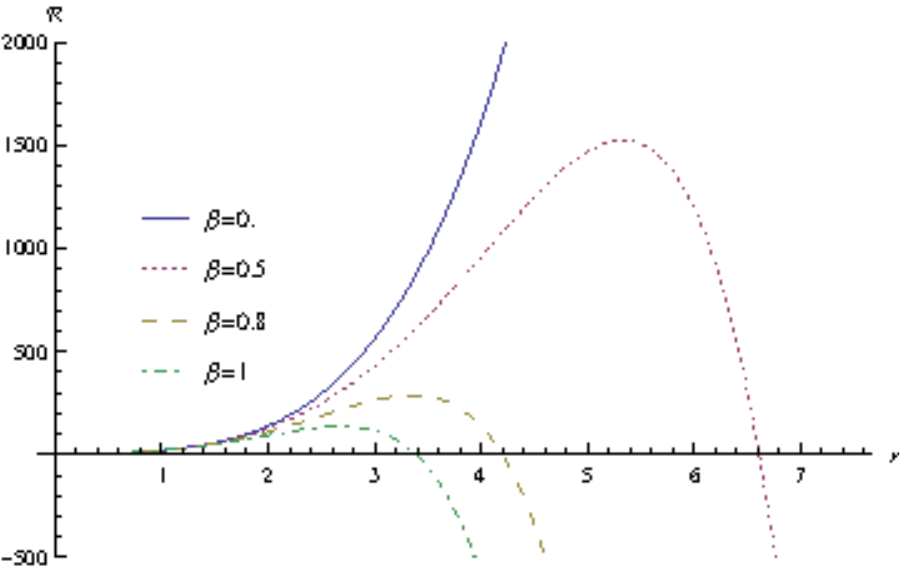}
\caption{\small     The allowed regions  for bounded orbits, $\mathcal{R} \ge 0$, are shown for different values of the nonlinear electromagnetic parameter $\beta$, from top to bottom $\beta=0, 0.5, 0.8, 1$; increasing $\beta$ decreases the allowed region $\mathcal{R} > 0$. The rest of the parameters are fixed as $m=1; \quad \tilde{E}=2.5; \quad \tilde{q}=0.1; \quad  \tilde{L}_{z}=0.5;  \quad \tilde{K}=4;  \quad a=0.9; \quad F_0=0.36; \quad  g_1=0.6. $ For $\beta=0$ there are no bounded orbits.
}
\label{figbeta}
\end{figure}


\subsection{$r$-motion}

The equation for the motion in $r$  is Eq. \eqref{rpunto}, $\dot{r}^2= {\mathcal{R}}(r),$  with ${\mathcal{R}}(r)$ given by, 
\begin{eqnarray}
\label{Radimension}
{\mathcal{R}} (r)&=& T(r)-  (r^2 + \tilde{K}) Q(r) \nonumber\\
&=&  \left[ \tilde{q} A_2(r) - a \tilde{L}_{z}  + \tilde{E} (r^2+a^2) \right]^2  - (r^2 + \tilde{K}) Q(r), \nonumber\\
\end{eqnarray}
and  the NLE contribution is through the parameter $\beta$ in  the electromagnetic potential $A_2(r)= g_1 r (1- \beta r^2)$
and  the metric function $Q(r)= \kappa F_0 (1- \beta r^2)^2/2 - 2 m r+r^2+a^2$; in what follows we  omit the tilde $\tilde{x}$ and fixed $\kappa =1$. 
The turning points, where $\dot{r}=0$, are given by the roots of $\mathcal{R}=0$ and  since $Q(r)$ is a fourth order polynomial,   $\mathcal{R}$  is a sixth order polynomial in ${r}$,  given by

\begin{eqnarray}
&&\mathcal{R}(r)=  \beta^2 (g_1^2 q^2-F_0) r^6  -2 \beta E g_1 q r^5
\nonumber\\
&&
+ \left[  E^2 -1 - \beta^2 F_0 K+ 2 \beta (F_0-g_1^2 q^2) \right] r^4 + \nonumber\\
&& \left[  2m+ 2 g_1 q (E+ a \beta (L_{z}-aE) \right] r^3 \nonumber\\
&&
+ \left[ g_1^2 q^2 + 2aE (aE - L_{z}) - K (1-2 \beta F_0) - a^2 -  F_0    \right] r^2  \nonumber\\
&& + \left( K 2m+2ag_1 q (aE-L_{z}) \right) r \nonumber\\
&&
+ a^2 (aE - L_{z})^2- K (a^2+F_0).
\label{R_pol6}
\end{eqnarray}

The previous  Eq.  (\ref{R_pol6})  with  $\beta=0$, making $F_0=Q_e^2$ and $g_1=Q_e,$ where $Q_e$ is the BH electric charge, we recover  the KN case

\begin{eqnarray}
\mathcal{R}^{KN}(r)&=&  ( E^2 -1 ) r^4 + 
\left[  2m+ 2 Q_e q E \right] r^3\nonumber\\
&&
+ \left[ Q_e^2 ( q^2 -1) + 2aE (aE - L_{z}) - K - a^2   \right] r^2 \nonumber\\
&& + \left(  K 2m +2a Q_e q (aE-L_{z}) \right) r \nonumber\\
&&
+ a^2 (aE - L_{z})^2  - K (a^2+ Q_e^2).
\end{eqnarray}
Which is a fourth degree polynomial.

In Fig. \ref{figZ} the generic behavior of the function  $\mathcal{R}$ is illustrated for the Kerr-Newman (KN) ($\beta=0$) and for the Stationary NLE BH ($\beta=0.5$).  In the KN BH   $\mathcal{R}$ is a fourth order polynomial then   $\mathcal{R}=0$ has  two real roots, one positive  and   one  negative and two complex conjugate roots.   One of the differences observed in   the Stationary NLE BH  is that,   for the same fixed parameters,    
$\mathcal{R}=0$ has  four real roots, two positive and two negative;
in addition of two complex conjugate roots. The two positive roots are the turning points and the  radius that  delimit the region of bounded orbits. This region is related to the BH capacity of growing an accretion disk. Therefore the Stationary NLE is a BH that will capture more easily test particles, this possibly leading to an accretion disk more rapidly increasing. One of the positive roots is in the interior of the BH, and the larger radius of the turning points defines the region  of the bounded orbits.
Geodesics of the KN spacetime are analyzed in \cite{Hackmann2013}, see also \cite{Cebeci2016}.

 In the next subsection we describe how Eq. (\ref{rpunto}) can  be integrated. 

\subsection{Solution in terms of hyperelliptic functions}

The radial coordinate in Eq. (\ref{rpunto}) can be integrated as

\begin{equation}
 \lambda = \int{\frac{dr}{\sqrt{\mathcal{R}(r)}}}.
\label{R_Eq}
\end{equation}

If we knew a zero of $\mathcal{R}(r_{0})=0$   (that might be a circular orbit radius or a turning point), then we could reduce the degree of $\mathcal{R}$ by substituting $r \mapsto 1/x + r_{0}$. The resulting fifth degree polynomial, 
$\mathcal{R}_x$, is given by

\begin{equation}
\mathcal{R}_x= \sum _{j=0}^{5} \frac{a_{j}}{a_{5}}x^{j}, \quad a_{j}=\frac{(\pm 1)^{j}}{(6-j)!} \frac{d^{(6-j)} \mathcal{R}}{dr^{(6-j)}} |_{r_{0}},
\end{equation}
this leads to  the hyperelliptic differential equation of  first kind,

\begin{equation}
x \frac{dx}{d \lambda}= E \sqrt{a_5 \mathcal{R}_x},
\end{equation}



The equations  involving hyperelliptic Riemann surfaces of genus 2 and one relevant degree of freedom are integrated in the framework of the Jacobi inversion problem, using a reduction to the   $\theta$-divisor (Jacobi Theta Function) on the Jacobi variety, i.e., to the set of zeros of the   $\theta$-function. The explicit solutions are given in terms of the Kleinian  $\sigma$-functions and their derivatives; therefore the solution for $r(\lambda)$ is given by

\begin{equation}
r( \lambda)= \mp \frac{\sigma_2}{\sigma_1} \left( 
\begin{array}{c}
f(\tau- \tau_{0}^{\prime})\\
\tau- \tau_{0}^{\prime}
\end{array}
 \right) + r_{0},
\end{equation}
where $\sigma_{i}, \quad i=1,2$  are the derivatives of the Kleinian $\sigma$-function and $f$ describes the $\theta-$divisor, i.e. $\sigma((f(z),z)^t)=0$.
The same procedure can be applied to the Eq. for the $\theta-$ motion that is also a sixth degree polynomial  in $x= \cos \theta.$ 
This method applied to  particle motion in General Relativity was  introduced in \cite{Enolskii2003}, \cite{Hackmann2008},  \cite{Hackmann2008b}. Therefore, in principle, the analytic solutions can be determined, however  in this contribution we will content
ourselves with a qualitative description of the possible test particle bounded orbits.

\subsection{Effective potential }

Eq.  \eqref{Radimension} is a quadratic equation in $E$, that can be written in terms of two  effective potentials  $V_{\rm eff \pm}$  as

\begin{eqnarray}
\label{R_eff_pot}
{\mathcal{R}} (r) &=& (E-V_{\rm eff +})(E- V_{\rm eff  -}),\nonumber\\
V_{\rm eff \pm} &=& \frac{1}{(r^2+a^2)} \left\{ (a {L}_{z}- {q} A_2(r)) \pm \sqrt{Q(r) \left( {K} + r^2  \right) } \right\}, \nonumber\\
\end{eqnarray}

 The positive square root is  the one that corresponds to a 4-momentum pointing toward the future. Since $V_{\rm eff -}$ can be negative then a region  exists from which energy can be extracted  by a Penrose process, that region  is called the effective ergoregion \cite{DenRuffini1974}; for stationary charged BH the energy extraction can be at the expense of electromagnetic energy or from rotating energy.
In Figs.  \ref{figEffPot1}- \ref{figEffPot2} are plotted the effective potentials  for $F_0$ positive and $F_0$ negative,   and for different values of $\beta$, respectively. The available regions for the motion of the test particle are the regions above and below the curves; it can be seen the presence of maxima and minima, that indicates there are circular orbits (where $V_{\rm eff}^{\prime}=0$  ) both unstable and stable.

\subsection{Circular orbits}

The spherical orbits are defined by $\mathcal{R}=0$ and  $\mathcal{R}^{\prime}=0$.
This last condition given by

\begin{eqnarray}
&&\mathcal{R}^{\prime} = 6 \beta^2 (g_1 q^2-F_0) r^5  - 10 \beta E g_1 q r^4 \nonumber\\
&&
+  4 \left[  E^2 -1 - \beta^2 F_0 {K}+ 2 \beta (F_0-g_1^2 q^2) \right] r^3 + \nonumber\\
&& 3 \left[  2m+ 2 g_1 q (E+ a \beta (L_{z}-aE) \right] r^2 \nonumber\\
&&
+ 2 \left[ g_1^2 q^2 + 2aE (aE - L_{z}) - {K} - a^2 - (1- 2 \beta {K} ) F_0    \right] r  \nonumber\\
&& + \left(  {K} 2m +2ag_1 q (aE-L_{z}) \right)  =0,
\end{eqnarray}

From the two conditions, $\mathcal{R}=0$ and  $\mathcal{R}^{\prime}=0$,  we can derive the condition for the circular orbits as

\begin{eqnarray}
&& 4(q A_{2}^{\prime} + 2 r E)^2 Q(r) \left( r^2 + {K} \right) \nonumber\\
&&
-  \left( 2r Q(r) + Q(r)^{\prime} \left( r^2 + {K}^2 \right) \right)^2=0,
\end{eqnarray}
where $f^{\prime} = df(r)/dr$.  Numerical solutions can be found for this equation, showing that indeed circular orbits can exist in the vicinity of this black hole. In Figs. \ref{figX} and \ref{figY}
are shown the radius of the circular orbits $r_{c}$ for fixed parameters and varying $\beta$ and $F_0$. Note that there  are two possible circular orbits for most of the ranges, indicating that unstable and stable circular orbits may occur, in agreement to the effective potentials shape. Moreover, the circular orbits of the Stationary NLE-BH  have greater radius
than the KN one.

\subsection{Bounded orbit  regions,  $\mathcal{R} \ge 0$}

For a qualitative description of the test particle motion we have explored
the  regions in which bounded orbits can occur,  $\mathcal{R} \ge 0$, by  varying the parameters. Moreover,  $\mathcal{R}= \dot{r}^2 = 0$  at the turning points and the effect of varying the parameters modifies the position of the turning points; if the turning point is nearer the horizon, then the region with  bounded orbits is smaller
and conversely.    In  Figs.\ref{figLz}-\ref{figbeta}  are explored the allowed regions    (only the regions of interest $r \ge 0$  are shown), first varying the test particle parameters $\tilde{L}_{z}$, $\tilde{E}$ and $\tilde{q}$;:

(i) Increasing $\tilde{L}_{z}$  the turning point   is nearer the horizon, then  the  allowed region for bounded orbits, $\mathcal{R} (r) \ge 0$ is smaller, see Fig.\ref{figLz}.

(ii) As $\tilde{E}$ increases  the  allowed region $\mathcal{R} (r)$ is enlarged  and the turning point is  farther from the horizon.
For $\tilde{E} <1$ there is not allowed motion in the vicinity of the black hole, i.e. $\mathcal{R} < 0$,  see Fig.\ref{figE}.
 In absence of positive real roots the orbits are of transit type: the particle starts at $\pm \infty$ comes to a point of closest approach, $r=r_o$ and then goes
 back to infinity.

(iii) As $\tilde{q}$ decreases  the  allowed region $\mathcal{R} (r)$ is enlarged  since the turning point is  farther from the horizon.
The maximum allowed region is for the uncharged test particle, $\tilde{q}=0$,  see Fig.\ref{figq}.

Now varying the black hole parameters  $a, Q, \beta$ we obtain the following behaviors:

(i) As $a$ increases   the  allowed region $\mathcal{R} (r)$ is enlarged  and the turning point  is farther from the horizon;
regions are not very different when varying  $a$; see Fig.\ref{figa}.

(ii)  Increasing the BH charge, $g_{1}=Q_e$, makes the turning point be  nearer the horizon and  the  allowed region $\mathcal{R} (r)$ becomes smaller.
 For the uncharged (Kerr) BH, $Q_e=0$ there are not  positive real roots, then the orbits are of transit type;  see Fig.\ref{figQ}.

(iii) As $\beta$ decreases  the  allowed region $\mathcal{R} (r)$ is enlarged  or the turning point is  farther from the horizon.
For the linear electromagnetic case, $\beta=0$ there are not  positive real roots, then the orbits are of transit type;  see Fig.\ref{figbeta}.

\subsection{{$\phi$}-motion}

We first note that the  possibility exists for  a  test particle coming with a certain angular momentum,   it   reverses its motion;  i.e. it may happen that $\dot{\phi}=0$, the equation that determine the corresponding $\theta$-angle is
\begin{equation}
P(x) Q(r)+ (a^2-x^2)T(r)=0,
\end{equation}
where $x=a \cos \theta$ that is a third degree polynomial. Otherwise the motion resembles the KN one.


\section{Birefringence}

To consider  the massless particles motion, we must take the limit $\mu=0$ in Eqs. (\ref{thetapart})-(\ref{rpart}). Recall that massless test particle trajectories are not the trajectories followed by light rays, since in NLE  these are governed by the null geodesics of an effective optical metric \cite{Pleban}, that is determined by the NLE Lagrangian.
In fact birefringence occurs in nonlinear electrodynamics with the exception of the Born-Infeld (BI) theory \cite{Pleban}; the propagation of signals in BI theory  was studied in \cite{Salazar89}. 

Given a Lagrangian depending on the two electromagnetic invariants $F$ and $G$, $L(F,G)$, there are two effective optical metrics $\gamma^{(1,2) \mu \nu}$ given in terms of derivatives of the Lagrangian respect $F$ and $G$, these are,

\begin{eqnarray}
\gamma^{(1) \mu \nu }&=&   \left( L_{F} - 2 L_{GG} F   \right) g^{\mu \nu} - 4 L_{GG} F^{\mu}_{. \lambda} F^{\lambda \nu}, \nonumber\\
\gamma^{(2) \mu \nu }&=&   L_{F}  g^{\mu \nu}   - 4 L_{FF} F^{\mu}_{. \lambda} F^{\lambda \nu}, 
\label{EffOpt}
\end{eqnarray}
where $g^{\mu \nu}$ is the spacetime metric, in our case the Stationary NLE-BH. For  Maxwell electrodynamics, $L(F)= F$,   $\gamma^{(1) \mu \nu}= \gamma^{(2) \mu \nu} = g^{\mu \nu}$ and birefringence does not occur. 
Therefore in the spacetime of the  Stationary NLE-BH birefringence will take place and we can determine the two effective optical metrics from Eqs. (\ref{EffOpt}) by means of the chain rule, since we know the derivatives of the Lagrangian respect to the coordinates $r$ and $\theta$; although  straightforward,  the procedure is cumbersome, for instance the expression for $L_{FF}$ becomes

\begin{eqnarray}
&& L_{FF} = 2 \frac{L_{,r \theta}}{F_{,r} F_{, \theta}} -\frac{F_{,r \theta}}{F_{,r} F_{, \theta}} \left[ \frac{L_{, \theta}}{ F_{, \theta}}+ \frac{L_{,r }}{F_{,r} } \right]  \nonumber\\
&& + \frac{1}{F_{,r}^2} \left[ L_{,r r} - \frac{L_{,r }F_{,rr}}{ F_{, r}} \right]
+ \frac{1}{(F_{,\theta})^2} \left[ L_{,\theta \theta} - \frac{L_{, \theta} F_{,\theta \theta}}{ F_{, \theta}}\right],\nonumber\\
\end{eqnarray}
analogously  the rest of the derivatives, $L_F$, $L_{GG}$,  can be determined since we do know the Lagrangian and the electromagnetic invariants as functions of $(r, \theta)$. 

Undoubtedly it would be interesting to determine the light trajectories in this metric, however we leave it for future research.



\section{Conclusions}

We have studied the properties of the stationary axisymmetric nonlinear electromagnetic spacetime that generalizes the Kerr-Newman BH. The solution possesses mass, rotation, electric and magnetic charges and three parameters associated to the NLE: $\beta$ of the electromagnetic potentials and $F_0$,$G_0$ related to the BH electric and magnetic charges.  From the analysis we conclude:

The sign of the nonlinear parameter $F_0$ determines the asymptotics of the spacetime: de Sitter if $F_0 <0$ and anti de Sitter when $F_0 >0$. While the value of $\beta$ determines the number of horizons, being then two critical values $\beta_1$ and $\beta_2$, for which there is only one horizon;
 if $\beta < \beta_1$  no horizons occur as well as if $\beta > \beta_2$; and for $\beta_1 < \beta < \beta_2$ the BH presents two horizons; in addition, if $F_{0} < 0$  there is the cosmological horizon.

 From the Kretschmann  invariant we see that the introduction of the NLE field affects the curvature. Assuming $\beta >0$ and $F_0 >0$ (in agreement with physically reasonable  energy conditions) then the Stationary NLE BH curvature is larger than the one of KN; and accordingly,   the bounded orbits are closer to the horizon for the rotating NLE BH than for KN.

For the motion of a charged test particle, the Hamilton-Jacobi equations turn out to be separable, likewise for the Kerr-Newman case, and in principle analytic solutions can be derived for the geodesic equations for $\theta$ and $r$,  related to a sixth degree polynomial; we did not follow that path and instead describe the regions allowed for the bounded motion of a charged test particle by varying the parameters; the allowed regions are illustrated in the plots for $\Theta( \theta) \ge 0$ and $\mathcal{R}(r) \ge 0$.
Among the effects of the introduction of the NLE field are the shrinking of  the regions allowed for  test particle bounded orbits;
 the allowed regions in terms of $\theta$  are defined by the positive real roots of $\Theta (\theta)=0$, and  there are three possible cases:
 (i) the first including  the whole range  $0 < \theta < \pi$, except one of the poles; (ii)  a strip defined by a $\theta_{\rm min} < \theta < \theta_{\rm max};$ (iii) the allowed region splits into two  disconnected regions one above and the other below the equatorial plane, in all cases one of the poles is unreachable:  the relative signs of the magnetic charge and the charge of the test particle determine which one of the poles is unreachable; if $\tilde{q} f_{1} >  0$ then test particle cannot reach $\theta=0$, while if  $\tilde{q} f_{1} < 0$ then $\theta= \pi$ is unreachable. 

For the radial motion circular orbits appear  in agreement with the shape of the effective potentials that presents maxima and minima. There is also the possibility of energy extraction since the effective potentials have regions of negative values.   In general the $r$-motion  is qualitatively  the same than in  KN spacetime but some differences arise: the NLE parameter modifies the number of bound orbits and  there exist a second circular orbit outside the horizon. 
The regions allowed for the test particle bounded orbits are closer to the horizon if the NLE parameter $\beta$ increases  and are larger for the uncharged BH (Kerr case). Augmenting the BH charge, the test particle charge or the nonlinear parameter $\beta$ results in smaller radius for the turning points that delimit the regions of bounded orbits, as a consequence,  the Stationary NLE BH may increase its accretion disk more easily.

The birefringence process takes place in this spacetime  and it is explained  how to obtain the effective optical metrics that determine light trajectories in this NLE spacetime.


\appendix


\section{Roots}\label{AppA}

The  metric  function representing a rotating charged black hole with nonlinear electrodynamics in  a Kerr-like spacetime  is

\begin{equation}
Q(r)=\frac{\kappa F_{0}}{2}\left(1-\beta r^2\right)^2-2mr+r^2+a^2,
\end{equation}

which we rewrite as follows as

\begin{equation}\label{QAi}
Q(r)=A_{1}r^4+A_{2}r^2+A_{3}r+A_{4},
\end{equation}

where $A_{i}$ are the coefficients defined as

\begin{eqnarray}
A_{1}&=&\frac{\kappa F_{0}\beta^2}{2},\quad A_{2}=1-\kappa F_{0}\beta, \nonumber\\
A_{3} &=& -2m,\quad A_{4}=\frac{\kappa F_{0}}{2}+a^2.
\end{eqnarray}

According to the Cardano-Ferrari method it is possible to determine analytical expressions for a fourth degree polynomial. The roots of (\ref{QAi}) are

\begin{equation}
r_{1,2}=\frac{1}{2}\sqrt{-\frac{2A_{2}}{3A_{1}}+\mathcal{B}}\mp\frac{1}{2}\sqrt{-\frac{4A_{2}}{3A_{1}}-\mathcal{B}-\frac{2A_{3}}{A_{1}\sqrt{-\frac{2A_{2}}{3A_{1}}+\mathcal{B}}}},
\end{equation}

and

\begin{equation}
r_{3,4}=-\frac{1}{2}\sqrt{-\frac{2A_{2}}{3A_{1}}+\mathcal{B}}\mp\frac{1}{2}\sqrt{-\frac{4A_{2}}{3A_{1}}-\mathcal{B}+\frac{2A_{3}}{A_{1}\sqrt{-\frac{2A_{2}}{3A_{1}}+\mathcal{B}}}},
\end{equation}

where $\mathcal{B}$ is defined as

\begin{equation}
\mathcal{B}=\frac{2^{1/3}\left(A_{2}^2+12A_{1}A_{4}\right)}{3A_{1}\mathcal{A}}+\frac{\mathcal{A}}{3(2)^{1/3}A_{1}},
\end{equation}

and $\mathcal{A}$ we define it as

\begin{eqnarray}
&& \mathcal{A}^{3} = 2A_{2}^3+27A_{1}A_{3}^2-72A_{1}A_{2}A_{4} \nonumber\\
&& +\sqrt{-4 \left(A_{2}^2+12A_{1}A_{4}\right)^3+\left(2A_{2}^3+27A_{1}A_{3}^2-72A_{1}A_{2}A_{4}\right)^2}.\nonumber\\
\end{eqnarray}


{\bf Acknowledgements}
N. B. acknowledges partial support by CONACYT-Project 284489.
GG-C acknowledges the support from the Consejo Nacional de Ciencia y Tecnología (CONACYT), Mexico, 
Estancias Posdoctorales por M\'exico 2021, grant No. 1031130.



\begin{thebibliography}{}
\bibitem{LIGO2016}
B. P. Abbot et al. [LIGO Scientific and Virgo Collaborations], {\sl Tests of general relativity with GW150914}, 
Phys. Rev. Lett.  {\bf 116}, 221101 (2016).

\bibitem{ALAV2021}
B. P. Abbot et al. [LIGO Scientific, Virgo Collaborationsand KAGRA Collaboration],
{\sl GWTC-3: Compact Binary Coalescences Observed by LIGO and Virgo During the Second Part of the Third Observing Run}
[arXiv: 2111.03606]

\bibitem{Garcia_Gustavo2019}
A. A. Garcia-Diaz and G. Gutierrez-Cano, {\sl Linear superposition of regular black hole solutions of Einstein nonlinear electrodynamics},
Phys. Rev. D {\bf 100}, 064068 (2019)

\bibitem{AGarcia2022}
 A. A. Garc\'ia-D\'iaz, {\sl Stationary Rotating Black Hole Exact Solution within Einstein-Nonlinear Electrodynamics,}
 {arXiv:2112.06302}
        
\bibitem{AGarcia2022B}
 A. A. Garc\'ia-D\'iaz,
 {\sl  AdS–dS Stationary Rotating Black Hole Exact Solution within Einstein–Nonlinear Electrodynamics},
Annals of Physics {\bf 441} 168880 (2022), 
[arXiv: 2201.10682].  Note that Eq. (90) for $Q(r)$ contains an extra factor $F_0$ in its first
term while in $\mu$ from Eq. (111)  the $\Lambda$ term is spurious.



\bibitem{SalazarGarciaPleb1987}
H. Salazar, A. Garcia, and J. F. Pleba\'nski, 
{\sl Duality rotations and type D solutions to Einstein equations with nonlinear electrodynamics sources},
J. Math. Phys. {\bf 28}
2171 (1987).


\bibitem{BornInfeld34}
M. Born and L. Infeld, {\sl Foundations of a new field theory},
Proc. Roy. Soc. {\bf A 144}, 125 (1934).

\bibitem{footnote}
{The roots of the polynomial are shown explicitly in Appendix \ref{AppA}}

\bibitem{Fayos2017}
R. Torres, F. Fayos, {\sl On regular rotating black holes},
Gen. relativ. Gravit. {\bf 49}:2 (2017) 

\bibitem{Gustavo2020}
A. A. Garcia-Diaz, G. Gutierrez-Cano, {\sl Regularity conditions for spherically symmetric solutions of Einstein-nonlinear electrodynamics equations}
Annals of Physics {\bf 422} 168323 (2020)

\bibitem{HawkingEllis}
        S. W. Hawking and G. F. E. Ellis,
        {\sl The large scale structure of space-time,}
        Cambridge University Press, 1973.

\bibitem{Carroll}
S. M. Carroll, {\sl Spacetime and Geometry}, AddisonWesley, 2004.

\bibitem{Carter68}
        B. Carter,
        {\sl Global Structure of the Kerr Family of Gravitational Fields,}
        Phys. Rev. \textbf{174}, (1968).

\bibitem{Visinescu2012}
M. Visinescu, {\sl Higher order first integrals, Killing tensors and Killing-Maxwell system}, J.  Phys. Conf.  Ser.  {\bf 343}  012126 (2012)

\bibitem{Mino2003}
Y. Mino, {\sl Perturbative approach to an orbital evolution around a supermassive black hole},
Phys. Rev. D {\bf 67}, 084027 (2003)

 \bibitem{Pleban}
S. Alarcon Gutierrez, A. L.  Dudley and J. F. Plebański,  {\sl Signals and discontinuities in general
relativistic non-linear electrodynamics} J. Math. Phys. {\bf 22} 2835–48 (1981)

\bibitem{Hackmann2013}
E. Hackmann and Hongxiao Xu;   {\sl Charged particle motion in Kerr-Newmann space-times},
Phys. Rev. D {\bf 87}, 124030 (2013)

\bibitem{Cebeci2016}
H. Cebeci,  N. Ozdemir, and  S. Sentorun,
{\sl Motion of the charged test particles in Kerr-Newman-Taub-NUT spacetime and analytical solutions},
Phys. Rev. D {\bf 93}, 104031 (2016)

\bibitem{Enolskii2003}
V. Enolskii, M. Pronine, and P. Richter, {\sl  Double Pendulum and $\theta$ -Divisor},  J. Nonlinear Sci.
{\bf 13}, 157 (2003).

\bibitem{Hackmann2008}
E. Hackmann and C. Laemmerzahl,
{\sl Geodesic equation in Schwarzschild-(anti-)de Sitter space-times:
Analytical solutions and applications}, Phys. Rev. D {\bf 78}, 024035 (2008)

\bibitem{Hackmann2008b}
E. Hackmann and  C. Laemerzahl,  {\sl Complete Analytic Solution of the Geodesic Equation in Schwarzschild–(Anti-)de Sitter
Spacetimes},   Phys. Rev. Lett.  {\bf 100}, 171101 (2008).

\bibitem{DenRuffini1974}
G. Denardo, L. Hively, R. Ruffini, {\sl On the generalized ergosphere of the Kerr-Newman geometry}, 
 Phys.  Lett. B  {\bf 50}, 270-272 (1974).

\bibitem{Salazar89}
H. Salazar Ibarguen, A. Garcia,  J. Pleba\'nski, {\sl Signals in nonlinear electrodynamics invariant under duality rotations}, J. Math. Phys {\bf 30}, 2689 (1989).



\end{thebibliography}
\end{document}